\documentclass[twocolumn,aps,showpacs]{revtex4} 

\usepackage{amsmath} 
\usepackage{graphicx}
\usepackage{dcolumn}
\usepackage{bm}

\newcommand{\nuc}[2]{$^{#1}${#2}} 

\begin{document}

\title{Nuclear collective excitations using correlated realistic interactions: 
        \\ the role of explicit RPA correlations} 

\author{P.~Papakonstantinou} 
\email[Email:]{panagiota.papakonstantinou@physik.tu-darmstadt.de} 
\affiliation{Institut f\"ur Kernphysik, 
Technische Universit\"at Darmstadt, 
Schlossgartenstr.~9, 
D-64289 Darmstadt, Germany} 

\author{R.~Roth} 
\affiliation{Institut f\"ur Kernphysik, 
Technische Universit\"at Darmstadt, 
Schlossgartenstr.~9, 
D-64289 Darmstadt, Germany} 

\author{N.~Paar}\thanks{On leave of absence from Physics Department, Faculty of Science, University of Zagreb, Croatia}  
\affiliation{Institut f\"ur Kernphysik, 
Technische Universit\"at Darmstadt, 
Schlossgartenstr.~9, 
D-64289 Darmstadt, Germany} 

\begin{abstract} 
We examine to which extent 
correlated realistic nucleon-nucleon interactions, derived within the 
Unitary Correlation Operator Method (UCOM), 
can describe nuclear collective motion in the framework of first-order random-phase approximation (RPA). 
To this end we employ 
the correlated Argonne V18 interaction 
in calculations within the so-called ``Extended" RPA (ERPA) 
and investigate the response of closed-shell nuclei. 
The ERPA is a renormalized RPA version which 
considers explicitly the depletion of the Fermi sea 
due to long-range correlations  
and thus allows us to examine how these  
affect the excitation spectra. It is found that the effect on the properties of 
giant resonances is rather small. 
Compared to the standard RPA, where excitations are built on top of 
the uncorrelated Hartree-Fock (HF) ground state, their centroid energies decrease by 
up to 1~MeV, approximately, in the 
isovector channel. The isoscalar response is less affected in general. 
Thus, the disagreement between our previous UCOM-based RPA calculations and the 
experimental data are to be attributed to other effects, 
mainly to a residual three-body force 
and higher-order configurations.
Ground-state properties obtained within the ERPA 
are compared with corresponding HF and perturbation-theory 
results and  
are discussed as well. The ERPA formalism 
is presented in detail.
\end{abstract} 

\date{\today}  


\maketitle

\section{Introduction} 
\label{Sintro} 

For the description 
of nuclear collective excitations and especially giant resonances 
throughout the nuclear chart, the theory of choice 
is the random phase approximation (RPA)  
 --- and its various siblings: the quasi-particle RPA for open-shell nuclei, 
the renormalized RPA, the 
second-order RPA (SRPA) and of course the relativistic versions. 
The most primitive applications of the usual RPA  
--- i.e., non-relativistic, first-order, for closed-shell nuclei ---  
consist typically of a 
ground-state description based on the Woods-Saxon potential  
and a phenomenological residual particle-hole ($ph$) interaction, e.g., 
of Landau-Migdal type. An important development in the 
field has been the construction of increasingly refined 
energy functionals and corresponding interactions, 
with those of Skyrme type dominating the 
landscape. These have allowed what is termed 
``self-consistent" RPA calculations: 
the residual $ph$ interaction is derived from 
the same interaction used in the Hartree-Fock (HF) description of the 
ground state. 
Self-consistency ensures that certain important formal properties of the 
RPA solutions hold exactly, including the 
separation of spurious states and the preservation of sum rules. 
Moreover, physical links between the bulk properties of nuclear matter and 
those of collective nuclear states can be established. 
The self-consistent RPA offers also enhanced predictive power: 
By starting from a more and more refined energy functional 
one wishes to make reliable extrapolations towards the unknown 
territories of the nuclear chart. 
Phenomenology enters this scheme 
in the form 
of the effective two-nucleon (NN) and three-nucleon (NNN) interaction.  
Contrary to the bare NN and NNN interactions, 
these are tailored for the use in mean-field type calculations such as HF and RPA. 
An ansatz is typically made for the effective NN(N) force 
and an appropriate fitting strategy has to follow. 

Highly accurate parameterizations of the bare NN force are 
nowadays available~\cite{WSS1995,Mac2001,SKT1994}. 
At the same time, attempts are made to derive the NN, NNN, and many-nucleon force 
starting from first principles, within chiral perturbation theory~\cite{EnM2003,EGM2005}. 
The possibility is currently being explored to combine mean-field theory 
with such realistic NN potentials. Two methods have been developed recently  
for ``taming" the NN interaction and thus making it usable within models like HF and RPA: 
the $V_{\text{low-}k}$ low-momentum interaction 
is derived within renormalization group theory~\cite{BKS2003} and  
the $V_{\text{UCOM}}$ correlated NN interaction is derived using the 
unitary correlation operator method (UCOM)~\cite{FNR1998,NeF2003,RNH2004}. 
Although constructed following different formalisms, the two potentials have similar 
low-momentum matrix elements.  In this work we focus on the UCOM potential. 

Within the UCOM, the major short-range correlations, 
induced by the strong repulsive core and the tensor part 
of the bare NN potential, are described by a 
state-independent 
unitary correlation operator. 
This can be used to introduce 
correlations into an uncorrelated many-body state or, alternatively, 
to perform a similarity transformation of an  
operator of interest.  
Applied to a realistic NN interaction, the method 
produces a ``correlated" interaction, 
$\text{V}_{\text{UCOM}}$, 
which can be used 
as a universal effective interaction, 
for calculations 
within simple Hilbert spaces. 

The aim of the UCOM is to treat explicitly only the state-independent 
short-range correlations; long-range correlations should be 
described by the model space. 
This tells us already 
that the UCOM-based HF is not enough, since a Slater-determinant 
wavefunction is unable to describe correlations. It was found indeed 
that, although bound nuclei were obtained using the $V_{\text{UCOM}}$ 
already at the HF level, the binding energies were underestimated 
by about 4~MeV per nucleon. 
Second-order perturbation theory, however, constitutes 
a tractable and adequate extension to the ``zero-order" description provided by HF, 
as far as nuclear binding energies are concerned~\cite{RPP2006}. 
 
In a recent publication~\cite{PPH2006} we employed the UCOM Hamiltonian in 
standard, self-consistent RPA calculations 
to study nuclear giant resonances. The ground state was described by the 
uncorrelated HF state, as usual. 
The main focus was on the isoscalar (IS) giant monopole resonance (GMR), 
the isovector (IV) giant dipole resonance (GDR), and the IS giant quadrupole resonance (GQR). 
Highly collective states were indeed obtained for various closed-shell nuclei ranging 
from \nuc{16}{O} to \nuc{208}{Pb}. We achieved a reasonable agreement with the 
experimental centroid energies of the IS GMR. By contrast, 
the energies of the IV GDR and the IS GQR 
were overestimated by several MeV.  

Obviously, the $V_{\text{UCOM}}$ is not a traditional effective interaction. 
Partly because no long-range correlations are (effectively) included in the UCOM, 
the corresponding nucleon effective mass in nuclear matter 
obtained in a HF calculation is very low 
(less than half the bare nucleon mass). 
This is confirmed by the HF results in finite nuclei: the single-particle 
level density was found too low. It is also manifested by the above-mentioned RPA results 
on the GQR and GDR centroids. 
From the preceding discussion it follows that, besides the possible important 
role of missing three-body terms in the Hamiltonian (see Sec.~\ref{ScorrH}), another source of our 
failure to describe nuclear collective states quantitatively can be  
the inadequacy of the RPA method to take into account residual long-range correlations. 

The standard RPA is based on two assumptions which hint at possible remedies. 
First of all, only one-particle-one-hole excitations are taken into account. 
One can include higher-order configurations, starting with two-particle-two-hole 
within SRPA. Given that an extended model space is of great importance 
when using the $V_{\text{UCOM}}$, it is imperative to examine the effect; 
work along this line is in progress. 
The second assumption is that one can approximate the 
true RPA ground state by the HF ground state. 
One argument follows from the fact that the RPA is the theory of small-amplitude vibrations 
around the HF ground state, 
therefore correlations should be small anyway for the use of RPA to be justified. 
It is not obvious that this assumption holds when the UCOM Hamiltonian 
is used, given the large correction to the HF binding energies 
due to second-order~\cite{RPP2006} and RPA~\cite{BPR2006} correlations. 
Therefore, in this work 
we wish to examine the effect of explicit RPA ground-state correlations 
on our results for nuclear collective states. 

To this end we use a renormalized version of the RPA, developed 
in Refs.~\cite{CPS1996,CGP1998,VKC2000}, 
where excited states are built on top of the true RPA ground state. 
It is formulated in the 
single-particle basis which diagonalizes the one-body density matrix, 
i.e., the natural-orbital basis, 
and its equations are solved iteratively.  
Following Ref.~\cite{VKC2000}, 
we will call it Extended RPA (ERPA). 
The term has been used in the literature to describe also higher-order RPA models 
(such as SRPA), but 
this is not the case here. 
Renormalized formulations of the RPA, 
such as the ERPA,  
consider explicitly 
the depletion of the Fermi sea in the 
ground state due to RPA correlations.  
Besides corrected excitation properties, 
the ERPA allows us to evaluate also corrected ground-state properties, 
namely single-particle energies and occupation numbers. 
It is derived using the 
number-operator method, which, 
contrary to the quasi-boson approximation, does not 
suffer from double-counting the second-order contributions~\cite{Row1968b,VKC2000}. 
 
This paper is organized as follows. 
In Sec.~\ref{ScorrH} we outline the UCOM method and introduce the 
correlated Hamiltonian. 
In Sec.~\ref{Serpa} we present the ERPA method 
and the formalism used to evaluate ground-state properties and 
transition strengths. 
In Sec.~\ref{Sresults} we present our results. 
In Sec.~\ref{Ssummary} we summarize and discuss future perspectives. 
 
\section{The correlated Hamiltonian} 
\label{ScorrH} 

In this work we employ a correlated interaction 
constructed within the unitary correlation operator method (UCOM), 
based on the Argonne V18~\cite{WSS1995} nucleon-nucleon (NN) interaction. 
Here we outline the basic principles of the UCOM 
scheme~\cite{FNR1998,NeF2003,RNH2004,RHP2005}. 
More detailed yet compact descriptions of the method can be found in Refs.~\cite{RPP2006,PPH2006}. 

The basic idea is the explicit treatment of the interaction-induced 
short-range central and tensor correlations. These are imprinted into an 
uncorrelated many-body state $|\Psi\rangle$ (e.g., a Slater determinant) 
through a state-independent unitary transformation 
defined by the unitary correlation operator $C$, resulting in a correlated state 
$|\tilde{\Psi}\rangle$,
\begin{equation} 
|\tilde{\Psi}\rangle = C |\Psi\rangle .  
\end{equation} 
The correlation operator $C$ is written as a product of unitary operators 
$C_{\Omega}$ and $C_{r}$ describing tensor and central correlations, respectively. 
Both are formulated as exponentials of a Hermitian generator, 
\begin{equation}
\label{eq:correlator}
  C = C_{\Omega} C_{r}
  = \exp [-\text{i} \sum_{i<j} g_{\Omega,ij} ]
    \exp [-\text{i} \sum_{i<j} g_{r,ij} ].
\end{equation}
The construction of the two-body generators $g_r$ and $g_{\Omega}$ follows the physical mechanisms 
by which the interaction induces central and tensor correlations. 
The short-range central correlations, caused by the repulsive core of the interaction, 
are introduced by a radial distance-dependent shift pushing nucleons apart from each other 
if they are within the range of the core. 
Tensor correlations between two nucleons are generated by a spatial shift perpendicular to the radial direction. 
For a given bare potential, 
the corresponding correlation functions are determined by an energy minimization in the two-body system for each 
$(S,T)$ channel. 

Matrix elements of an operator $O$ 
with correlated many-body states 
$|\tilde{\Psi}\rangle$  
can be equivalently written as matrix elements of a ``correlated" (transformed)  
operator $\tilde{O}=C^{\dagger}OC$ and uncorrelated many-body states 
$|\Psi\rangle$, 
thanks to the unitarity of the correlation operator. 
Thus, one can work in simple Hilbert spaces (simple states) using correlated operators, 
as well as with bare operators and explicitly correlated states. 
By applying the unitary transformation to a bare NN interaction, 
a phase-shift equivalent correlated interaction is obtained, which is suitable for use in 
tractable model spaces~\cite{RNH2004,RHP2005,RPP2006}. 
The same transformation can then be applied to any 
other operator under study, as is needed for a consistent UCOM treatment. 

There are a couple of important issues arising in actual applications. 
In an $A$-body system the correlated operator contains irreducible 
contributions to all particle numbers. 
Within a cluster expansion of the correlated operator
\begin{equation} 
  \tilde{O} = C^{\dagger} O C = \tilde{O}^{[1]} + \tilde{O}^{[2]} + \cdots + \tilde{O}^{[A]}, 
\label{cexpansion}
\end{equation} 
where $\tilde{O}^{[n]}$ denotes the irreducible $n$-body contribution, we usually employ 
a two-body approximation, i.e., three-body and higher-order terms of the expansion are neglected. 
Starting from the uncorrelated Hamiltonian $H$ for the $A$-body system,
consisting of the kinetic energy operator $T$ and a two-body potential $V$, 
the formalism of the UCOM is used to construct 
the correlated Hamiltonian in two-body approximation
\begin{equation}
  H^{C2} = {\tilde{T}}^{[1]} + {\tilde{T}}^{[2]} + {\tilde{V}}^{[2]}
  = T + V_{\text{UCOM}},
\end{equation}
where the one-body contribution comes only from the uncorrelated kinetic energy ${\tilde{T}}^{[1]} = T$.
Two-body contributions arise from the correlated kinetic energy ${\tilde{T}}^{[2]}$ and the correlated 
potential ${\tilde{V}}^{[2]}$, which together constitute the phase-shift equivalent correlated interaction $V_{\text{UCOM}}$.

It has been verified that higher-order contributions due to short-range central correlations 
can be neglected in the description of nuclear structure properties \cite{RNH2004}. 
The tensor interaction, on the other hand, 
is long-ranged and thus generates long-range correlations 
in an isolated two-nucleon system. 
However, the long-range tensor 
correlations between two nucleons embedded in a many-nucleon system 
are suppressed by the presence of other nucleons, 
leading to a screening of the tensor correlations at large interparticle distances. 
In terms of the cluster expansion this screening appears through significant higher-order contributions. 
In order to avoid large higher-order contributions, 
and at the same time to effectively describe the screening effect, 
the range of the tensor correlation function 
--- more precisely, 
the ``correlation volume" $I_{\vartheta}^{(S,T)}$~\cite{RHP2005} ---  
is restricted during the parameterization procedure. 
Restricting the range of the tensor correlator has another important function, 
namely to ensure that only 
state-independent, short-range correlations are described by the UCOM. 
By varying the correlation volumes --- the only parameters entering the formalism ---  
a family of correlators and respective correlated 
interactions are obtained. 

The question is then how to optimize these parameters in order to 
best describe the screening effect and the separation of the two types 
of correlations. 
As demonstrated in Ref.~\cite{RHP2005}, 
this can be done 
with the help of exact few-body calculations. 
In particular, applications within the no-core shell model showed  
that for the Argonne V18 potential the value $I_{\vartheta}^{(1,0)}=0.09$~fm$^3$ 
leads to the best description of binding energies 
in \nuc{3}{H} and \nuc{4}{He}. 
For this choice of tensor correlator range the 
missing genuine three-nucleon interaction and  the ommitted higher-order 
terms of  the cluster expansion of the correlated Hamiltonian effectively cancel 
each other. 
As was subsequently shown within many-body perturbation theory~\cite{RPP2006}, 
and verified by RPA calculations~\cite{BPR2006}, this cancelation remains at work 
throughout the nuclear chart, as far as the binding energy is concerned. 
The same does not necessarily hold, however, for other ground-state observables, such as charge radii, 
or for excited-state properties. 

In this work we will use the correlated Argonne V18 potential with 
$I_{\vartheta}^{(1,0)}=0.09$~fm$^3$. 
No tensor correlator is employed in the triplet-odd channel, 
where the tensor interaction is much weaker. 
We start from a Hamiltonian which consists of the intrinsic 
kinetic energy and the $V_{\text{UCOM}}$ interaction 
derived from the Argonne V18 potential including the Coulomb potential, 
\begin{equation} 
\label{eq:hintr}
  \mathcal{H} = {H}_{\text{int}} 
  = T - T_{\text{cm}} + V_{\text{UCOM}} 
  = T_{\text{int}} + V_{\text{UCOM}} \;,
\end{equation} 
in two-body approximation. The intrinsic kinetic energy operator reads,
\begin{equation} 
\label{intrke}
  T_{\text{int}} 
  = T - T_{\text{cm}} 
  = \frac{2}{A} \frac{1}{m} \sum_{i<j}^{A} (\bm{p}_i - \bm{p}_j)^2 , 
\end{equation} 
where 
we assume equal proton and neutron masses, $m$. 
It is the two-body Hamiltonian $H_{\text{int}}$ of Eq.~(\ref{eq:hintr}) 
that has been used in Hartree-Fock (HF), perturbation-theory, and RPA calculations in Refs.~\cite{RPP2006,PPH2006} 
and that will be employed in this work too. 
In practice, two-body matrix elements in a harmonic-oscillator basis are the input 
to such calculations.  
The harmonic-oscillator basis is typically restricted to $13$ major shells, which warrants 
complete convergence of the HF results. Calculations with larger basis sizes are possible 
but increasingly time-consuming. 
The optimal value of the harmonic oscillator parameter is determined from an explicit energy 
minimization for different regions in the nuclide chart.

\section{The ERPA method} 
\label{Serpa} 

The method used in this work is a first-order RPA. 
The excited nuclear states are described as 
linear combinations of particle-hole ($ph$) and hole-particle ($hp$) configurations 
and created by a set of vibration creation operators. The RPA ground state 
(reference state) is the vacuum of the corresponding annihilation operators and contains 
induced long-range correlations. 
As mentioned in Sec.~\ref{Sintro}, in order to facilitate the actual solution of the RPA 
equations,  one usually approximates the reference state by the 
uncorrelated HF ground state when calculating the RPA matrix elements and the 
transition matrix elements. 
Here we use a renormalized formulation of the RPA that 
avoids this approximation and allows us not only to assess the 
effect of ground-state correlations on the excitation spectra, 
but also to evaluate some properties of the correlated RPA ground state. 
Following Ref.~\cite{VKC2000}, 
we call this formulation Extended RPA (ERPA) to 
distinguish it from the one 
which is based on the 
uncorrelated HF ground state. 
 
Next, 
we review the ERPA method based on Refs.~\cite{CPS1996,CGP1998,VKC2000} 
and we present the formalism used to evaluate ground-state properties and 
transition strengths. 
Expressions with explicit angular-momentum coupling are shown. 
Closed-shell nuclei are considered and spherical symmetry is assumed. 
The symbol $p(h)$ will represent all the quantum numbers of a particle 
(hole) state except the magnetic quantum number $m_p (m_h)$, 
i.e.,  the set of quantum numbers $\{n_{p(h)} \ell_{p(h)} j_{p(h)} t_{p(h)} \}$ 
of the nodes, orbital angular momentum, total angular momentum, 
and isospin. 
The letters $\alpha , \beta , ... $  
will be used to denote single particle states of either kind ($p$ or $h$).  
As a single-particle basis, the one that diagonalizes the 
one-body density matrix (OBDM) is chosen --- i.e., the natural-orbital basis. 
The ERPA formalism is based on this choice.

\subsection{The ERPA equations} 

The creation operator 
$Q^{\dagger}_{\nu ,{J}^{\pi}{M}}$ 
of an excited 
state $|\nu\rangle$ with multipolarity ${J}$  
and parity $\pi $ 
is written as a linear combination of 
renormalized $ph$ creation and 
annihilation operators 
$\mathcal{A}_{ph}^{{JM} \dagger}$,  $\mathcal{A}_{ph}^{{JM}}$, 
\begin{eqnarray} 
|\nu\rangle &=& Q^{\dagger}_{\nu ,{J}^{\pi}{M}}| 0 \rangle 
\nonumber \\ 
            &=& \sum_{ph}  
[ X_{ph}^{\nu , {J}^{\pi}           } \!\mathcal{A}_{ph}^{{JM} \dagger}  
- Y_{ph}^{\nu , {J}^{\pi}           } \!(-1)^{{J+M}} \!\mathcal{A}_{ph}^{{J-M}}] | 0 \rangle , 
\label{E:nu} 
\end{eqnarray} 
where $| 0 \rangle $ is the ground state, for which  
\begin{equation}
 Q_{\nu ,{J}^{\pi}{M}}| 0 \rangle = 0,  
\end{equation} 
and the sum runs over $ph$ pairs which can couple to 
total spin and parity ${J}^{\pi}$. 
The renormalized operators are 
written as linear combinations of $ph$ operators, 
\begin{equation}
\mathcal{A}_{ph}^{{JM} \dagger} = 
\sum_{n_p'n_h'} N_{n_pn_h,n_p'n_h'}^{\ell_pj_pt_p;\ell_hj_ht_h} 
[a_{n_p'\ell_pj_pt_p}^{\dagger} a_{n_h'\ell_hj_ht_h}]^{{JM}} 
, 
\label{E:Aop} 
\end{equation}
where the angular-momentum coupled operators are given by 
\begin{equation}
[a_{p}^{\dagger} a_{h}]^{{JM}} = \sum_{m_pm_h} 
\langle j_p m_p j_h m_h | {JM}\rangle 
(-1)^{j_h+m_h} 
a_{pm_p}^{\dagger} a_{h-m_h} .
\end{equation}
The choice of the single particle basis and the requirement that 
\begin{equation}
\langle 0 | 
[ \mathcal{A}_{p'h'}^{{JM} }, 
  \mathcal{A}_{ph}^{{JM} \dagger}  ] 
| 0 \rangle = \delta_{pp'}\delta_{hh'} 
 , 
\end{equation}
following from the fermionic character of the 
operators, dictate that the matrix $N$ be given by~\cite{CGP1998}  
\begin{eqnarray} 
N_{n_pn_h,n_p'n_h'}^{\ell_pj_pt_p;\ell_hj_ht_h} 
&=& 
\delta_{n_pn_p'}  
\delta_{n_hn_h'} 
(\rho_h - \rho_p)^{-1/2} 
\nonumber \\ 
&\equiv&  
\delta_{n_pn_p'}  
\delta_{n_hn_h'} 
D_{ph}^{-1/2}  .  
\label{E:Nmat} 
\end{eqnarray}
Here 
\begin{equation}
\rho_{\alpha} = \langle 0 | a^{\dagger}_{\alpha m_{\alpha}} a_{\alpha m_{\alpha}} 
| 0 \rangle  
\end{equation}
denotes the occupation probability of the single-particle state 
$\alpha$, which is independent of $m_{\alpha}$. 
It can be shown that, under the above circumstances, 
the amplitudes $X$ and $Y$ 
of Eq.~(\ref{E:nu}) obey the orthonormalization condition 
\begin{equation} 
\sum_{ph} ( 
X_{ph}^{\nu ' , {J}^{\pi}            \ast} X_{ph}^{\nu , {J}^{\pi}           } 
- 
Y_{ph}^{\nu ' , {J}^{\pi}            \ast} Y_{ph}^{\nu , {J}^{\pi}           } 
) 
=\delta_{\nu\nu '} 
. 
\end{equation} 

The equations determining the 
amplitudes $X$ and $Y$ 
are obtained 
by using the equations-of-motion method~\cite{Row1968a}. 
As usual, they are written as a matrix equation in the $ph$ space, 
\begin{equation} 
\left( 
\begin{array}{cc}  
A & B \\ -B^{\ast} & -A^{\ast} 
\end{array} 
\right) 
\left( 
\begin{array}{c} 
X^{\nu} 
\\ 
Y^{\nu} 
\end{array} 
\right) 
= E_{\nu}  
\left( 
\begin{array}{c} 
X^{\nu} 
\\ 
Y^{\nu} 
\end{array} 
\right) 
, 
\label{E:ERPAeq}  
\end{equation}  
where $E_{\nu}$ is the excitation energy of the state $|\nu\rangle $ 
and the quantum numbers ${J}^{\pi}$ are implicit.  
The elements of the matrices $A$ and $B$, 
which are independent of ${M}$, are given by 
\begin{equation}
\label{E:Agen} 
A_{php'h'}^{{J}} = 
\langle 0 | [\mathcal{A}_{ph}^{{JM}}, \mathcal{H} ,\mathcal{A}_{p'h'}^{{JM} \dagger }] | 0 \rangle  
\end{equation}
and 
\begin{equation}
\label{E:Bgen} 
B_{php'h'}^{{J}} = 
-\langle 0 | [\mathcal{A}_{ph}^{{JM}}, \mathcal{H} ,(-1)^{{J}+{M}} 
\mathcal{A}_{p'h'}^{{J-M}}] | 0 \rangle  
, 
\end{equation} 
where $\mathcal{H}$ is the Hamiltonian of the system. 
The symmetrized double commutators are defined as 
\[ 
[A,B,C] = \frac{1}{2} ( [A,[B,C] ] + [ [A,B], C] ) 
. 
\] 
Here we consider the two-body Hamiltonian, 
\begin{equation} 
\mathcal{H} = H_{\text{int}} = \frac{1}{4} \sum_{\alpha\beta \alpha ' \beta ' } 
H_{\alpha\beta\alpha'\beta'} 
a_{\alpha}^{\dagger}a_{\beta}^{\dagger} 
a_{\beta'}a_{\alpha'} 
,
\label{E:ham} 
\end{equation}  
containing a two-nucleon interaction 
and the $A$-dependent intrinsic part of the kinetic energy, 
as introduced in Sec.~\ref{ScorrH}, Eq.~(\ref{eq:hintr}). 

The remaining task is to evaluate the expectation values of the 
double commutators 
in Eqs.~(\ref{E:Agen}) and (\ref{E:Bgen}). 
In general, elements of both the one-body and the 
two-body density matrices will appear in the 
commutator of the Hamiltonian $H_{\text{int}}$ with a $ph$ operator. 
The elements of the two-body density matrix 
can be tackled using a linearization procedure~\cite{Bro1967,Row1970}, 
by which they are contracted with respect to a reference state (the ground state) 
so that a quantity linear in the $ph$ operators is obtained. 
Then, by substituting 
in Eqs.~(\ref{E:Agen}) and (\ref{E:Bgen}), 
the matrix elements of $A$ and $B$ can be expressed in terms 
of the OBDM only. 
In the standard RPA, one uses the HF state 
(for which $\rho_h=1$, $\rho_p=0$) 
as the reference state. 
In the present ERPA the correlated ground state $|0\rangle$ is 
chosen as the reference state. 
By performing this linearization, one obtains 
\begin{eqnarray} 
A_{php'h'}^{{J}} 
&=& \frac{1}{2}( 
D^{1/2}_{ph} D^{-1/2}_{p'h'} + 
D^{-1/2}_{ph} D^{1/2}_{p'h'}) 
\nonumber \\ 
& & \,\,\,\, \times  
(\delta_{hh'} \varepsilon_{p'p} - \delta_{pp'} \varepsilon_{hh'} )  
\nonumber \\ 
&& 
+ 
D^{1/2}_{ph} D^{1/2}_{p'h'} 
\langle p h^{-1} ;{J}  |  H_{\text{int}}  |  p'{h'}^{-1} ; {J} \rangle  
\label{E:Amat} 
\\ 
B_{php'h'}^{{J}} 
&=& 
[D_{ph} D_{p'h'} 
(1+\delta_{hh'})(1+\delta_{pp'})]^{1/2} 
\nonumber \\ 
& & \,\,\,\, \times  
\langle (p h^{-1} ;{J}) (p'{h'}^{-1} ; {J}^{-1})  |  H_{\text{int}}  |  0 \rangle  
\label{E:Bmat} 
\end{eqnarray}  
where $ | \alpha \beta ; J \rangle $ 
are the normalized and antisymmetrized, angular-momentum--coupled two-body states  
and 
\begin{equation} 
\varepsilon_{\alpha \alpha '} \equiv 
\sum_{\beta ,J} \rho_{\beta}  
 [(1+\delta_{\alpha\beta})(1+\delta_{\alpha '\beta})]^{1/2} 
\langle \alpha \beta ;J  |  H_{\text{int}}  |  \alpha' \beta  ; J \rangle  
\label{E:eaa} 
\end{equation} 
are matrix elements of the single-particle Hamiltonian (see Sec. \ref{S:gsp}). 
The quantity  $D_{ph}$ has been defined in Eq.~(\ref{E:Nmat}). 
For $\rho_h=1$, $\rho_p=0$ the usual RPA equations are recovered. 

According to Eqs.~(\ref{E:Amat}) and (\ref{E:Bmat}), the ERPA matrix (\ref{E:ERPAeq}) 
depends on the occupation probabilities $\rho_{\alpha}$. 
These, as we will see, depend on the  
forward and backward amplitudes $X$ and $Y$ 
of all the ${J}^{\pi}$ states of the nucleus, 
i.e., on the solutions of the ERPA equations. 
Therefore, the equations must be solved 
iteratively. Let us call this procedure the ERPA iteration. 
The occupation probabilities of the single-particle states can be 
expressed in terms of the 
amplitudes $X$ and $Y$ 
and the matrix $N$ of Eq.~(\ref{E:Nmat}), 
using the 
number-operator method~\cite{CPS1996}. 
Contrary to the quasi-boson approximation, this method does not 
suffer from double-counting the second-order contributions~\cite{Row1968b,VKC2000}. 
For the OBDM in a generic basis one finds~\cite{CGP1998,VKC2000}
\begin{eqnarray} 
\lefteqn{\langle 0 | a_{pm_p}^{\dagger} a_{p'm_{p'}} | 0 \rangle  
= } \nonumber \\ 
&&\sum_{hm_h \nu \nu'} 
(\delta_{\nu\nu'} 
 -\frac{1}{2} \! \sum_{p_1h_1} 
\langle 0   | a_{h_1m_{h_1}}^{\dagger} a_{p_1m_{p_1}} | \nu' \rangle  
\nonumber \\ && 
\mbox{} \hspace{3cm} \times 
\langle \nu | a_{p_1m_{p_1}}^{\dagger} a_{h_1m_{h_1}} | 0    \rangle 
) 
\nonumber 
\\ &&  
\mbox{}\hspace{8mm}\times 
\langle 0   | a_{p  m_p    }^{\dagger} a_{h  m_h    } | \nu  \rangle 
\langle \nu'| a_{h  m_h    }^{\dagger} a_{p' m_{p' }} | 0    \rangle 
, 
\label{E:gOBDMp} 
\\ 
\lefteqn{\langle 0 | a_{hm_h}^{\dagger} a_{h'm_{h'}} | 0 \rangle 
=  
\delta_{h,h'}\delta_{m_hm_{h'}} }
\nonumber \\ 
&-& 
\sum_{pm_p \nu \nu'} 
(\delta_{\nu\nu'} 
 -\frac{1}{2} \! \sum_{p_1h_1} 
\langle 0   | a_{h_1m_{h_1}}^{\dagger} a_{p_1m_{p_1}} | \nu' \rangle 
\nonumber \\ && 
\mbox{} \hspace{3cm} \times 
\langle \nu | a_{p_1m_{p_1}}^{\dagger} a_{h_1m_{h_1}} | 0    \rangle 
) 
\nonumber 
\\ 
&& 
\mbox{}\hspace{8mm}\times 
\langle 0   | a_{p  m_p    }^{\dagger} a_{h  m_h    } | \nu  \rangle 
\langle \nu'| a_{h  m_h    }^{\dagger} a_{p' m_{p' }} | 0    \rangle 
. 
\label{E:gOBDMh} 
\end{eqnarray} 
In the natural-orbital basis, and using angular momentum coupling, we have 
\begin{widetext} 
\begin{eqnarray} 
\langle 0 | a_{p}^{\dagger} a_{p'} | 0 \rangle 
&=& 
\hat{j}_p^{-2}  
\sum_{{J}^{\pi} \nu \nu' h }
\hat{J}^2       
[ 
\delta_{\nu\nu'} - \frac{1}{2} \sum_{p_1h_1} D_{p_1h_1}X_{p_1h_1}^{\nu',{J}^{\pi}} 
                                                       X_{p_1h_1}^{\nu ,{J}^{\pi} \ast}  
] 
D_{ph}^{1/2}D_{p'h}^{1/2}Y_{ph}^{\nu ,{J}^{\pi}} Y_{p'h}^{\nu',{J}^{\pi}\ast } 
, 
\label{E:OBDMp} 
\\ 
\langle 0 | a_{h}^{\dagger} a_{h'} | 0 \rangle 
&=& 
\delta_{hh'} -  
\hat{j}_h^{-2}    
\sum_{{J}^{\pi} \nu \nu' p  }
\hat{J}^2          
[ 
\delta_{\nu\nu'} 
- \frac{1}{2} \sum_{p_1h_1} D_{p_1h_1}X_{p_1h_1}^{\nu',{J}^{\pi}} 
                                                       X_{p_1h_1}^{\nu ,{J}^{\pi} \ast}  
] 
D_{ph}^{1/2}D_{ph'}^{1/2}Y_{ph}^{\nu ,{J}^{\pi}} Y_{ph'}^{\nu',{J}^{\pi}\ast } 
, 
\label{E:OBDMh} 
\end{eqnarray} 
\end{widetext} 
where $\hat{j}\equiv \sqrt{(2j+1)}$. %
These equations, which are exact up to terms $O(|Y|^4)$, are not linear in the OBDM elements. 
Therefore, for given $X$ and $Y$ amplitudes (at a given step of the ERPA iteration), 
they have to be solved iteratively as well: 
one begins with an initial guess for the occupation probabilities on the 
r.h.s., calculates the OBDM on the l.h.s., diagonalizes it, thus obtaining a new 
set of occupation probabilities, and so on, until convergence is reached. 
Let us call this the OBDM iteration. 

This double iterative procedure of solving the ERPA equations can be outlined as follows: 
In order to begin the ERPA iteration, we need an initial choice for 
the ground state. We start with the HF ground state. 
We build the ERPA matrix 
for each ${J}^{\pi}$ that we wish to take into account   
by using Eqs.~(\ref{E:Amat}) and (\ref{E:Bmat}) 
and assuming the HF occupation probabilities (1 for holes and 0 for particles). 
Then we solve the eigenvalue problem 
of Eq.~(\ref{E:ERPAeq}). 
At this stage, we have simply solved conventional RPA equations. 
Then we calculate a ``new" OBDM by using 
Eqs.~(\ref{E:OBDMp}) and (\ref{E:OBDMh}) and, as input, the $X$ and $Y$ amplitudes and 
the HF occupation probabilities.  
We proceed with a OBDM iteration and obtain new occupation probabilities, which 
hereafter replace the initial ones (the HF ones in this first step) and  
are used as input to build again the ERPA matrix for each ${J}^{\pi}$. 
Here begins the second ERPA-iteration step. We proceed as before, 
until convergence is reached. 

\subsection{Ground-state properties in the ERPA method} 
\label{S:gsp} 

The ERPA method provides information on the correlated RPA ground state. 
As we have seen, the natural orbitals can be determined and 
their occupation probabilities 
can be calculated up to fourth order in the 
backward amplitude. 
As a convenient way to quantify the ground-state correlations 
we will use the mean square deviation per particle  
of the OBDM~\cite{AHP1993}, defined by  
\begin{equation} 
\sigma=\text{Tr}[(\varrho - \varrho^{(0)})^2] / A  
.
\label{E:sigma} 
\end{equation} 
This quantity characterizes the deviation of the OBDM $\varrho$ of the 
correlated ground state from the OBDM $\varrho^{(0)}$ of the uncorrelated 
state described by a Slater determinant $|\Psi^{(0)}\rangle$. 
It describes how well $|\Psi^{(0)}\rangle$ 
approximates the correlated state. 
It is 
minimal when evaluated in the natural-orbital basis, 
with $|\Psi^{(0)}\rangle$ built of the 
$A$ hole states~\cite{Kob1969}. Then 
\begin{equation} 
\sigma = \sigma_{\min}=\frac{1}{A} 
\left[ 
\sum_h \hat{j}_h^2 
(1-\rho_h)^2 
+ \sum_p \hat{j}_p^2   
\rho_p^2 
\right] 
\label{E:sigmaNO} 
\end{equation} 
expresses the depletion of the hole states and the 
occupation of the particle states as an average. 
In HF theory $\sigma_{\min}$ obviously vanishes, as does the Fermi-sea depletion, 
while in realistic cases $\sigma_{\min}$ is expected to take values of the order 
$10^{-2}$~\cite{AHP1993,JMN1987}, corresponding to a Fermi-sea depletion of at least 10\%.  
It has long been observed~\cite{JMN1987} that values way below $10^{-2}$ indicate that 
short- and medium-range correlations have not been (adequately) taken into account 
by the model in question. This is typically the case, e.g., in RPA calculations.  
We will return to this point in Sec.~\ref{Sresults}.  

Removal and pick-up centroid energies, $\varepsilon^{(-)}$ 
and $\varepsilon^{(+)}$, 
respectively, can also be 
defined~\cite{Row1968a,Row1970,Fri1975} and evaluated within the ERPA. 
The removal energy $\varepsilon^{(-)}_h$ corresponding to the hole state $h$ is the 
centroid of the distribution 
\begin{eqnarray}  
S^{(-)}_h (E) &=& \sum_{fm_h} |\langle f | a_{hm_h} | 0 \rangle |^2 \delta (E-E_f)  
  \nonumber \\ &=& \langle 0 | a_{hm_h}^{\dagger} \delta (E-H_{\text{int}}) a_{hm_h} | 0 \rangle , 
\end{eqnarray} 
where $ | f \rangle $, $E_f$ are the eigenstates and energies of the 
$(A-1)$-system, $H_{\text{int}} |f\rangle = E_f|f\rangle$. 
One finds, using again a linearization procedure, 
\begin{equation} 
\varepsilon^{(-)}_{h} = -\sum_{\alpha, J} \rho_{\alpha} 
(1+\delta_{\alpha h}) 
\langle \alpha h ; J | H_{\text{int}} | \alpha h ; J \rangle \equiv - \varepsilon_h 
. 
\label{E:spvareh} 
\end{equation} 
Similarly one finds the centroid of the energy corresponding to picking up a particle 
at the state $p$, 
\begin{equation} 
\varepsilon^{(+)}_{p} =  \sum_{\alpha, J} \rho_{\alpha} 
(1+\delta_{\alpha p}) 
\langle \alpha p ; J | H_{\text{int}} | \alpha p ; J \rangle \equiv  \varepsilon_p 
. 
\label{E:spvarep} 
\end{equation} 
Defined in the natural-orbital basis, 
$\varepsilon_h$ and 
$\varepsilon_p$ 
no longer correspond to the eigenvalues of the single-particle (s.p.)  
Hamiltonian --- in contrast to the HF case. 
A s.p. Hamiltonian $H_{\text{int}} ^{(1)}$ can be defined, 
in principle, 
as the one-body term of the nuclear Hamiltonian operator, Eq.~(\ref{E:ham}), when written in a 
normal-ordered form~\cite{Row1970,Hey1990}.
It differs from the HF s.p. Hamiltonian only by a similarity transformation 
and therefore posesses the same eigenvalues. 
If, however, one {linearizes} the two-body part of the normal-ordered 
total Hamiltonian $H_{\text{int}}$ and adds the thus obtained one-body operator 
to the genuine one-body part $H_{\text{int}}^{(1)}$, one 
obtains a s.p. Hamiltonian $H_{\text{int}} ^{\text{s.p.}}$ whose  
matrix elements in the natural-orbital basis are given by Eq.~(\ref{E:eaa}). 
Its diagonal matrix elements coincide with the 
$\varepsilon_h$, $\varepsilon_p$ energies 
defined in Eqs.~(\ref{E:spvareh}), (\ref{E:spvarep}).  
It is possible to diagonalize the s.p. Hamiltonian in order to obtain its 
eigenvalues, $e_{\alpha}$. 

All s.p. energies defined above have been formally evaluated in terms 
of matrix elements of the two-body Hamiltonian $H_{\text{int}} $. They should be further corrected 
for the center-of-mass energy, as was done in Ref.~\cite{PPH2006} for the 
HF s.p. energies. For the purposes of this work, however, 
it will suffice 
to  
compare our 
uncorrected results with the uncorrected HF energies, 
in order to show the relative effect of RPA correlations. 

One should keep in mind, that only for s.p. levels around 
the Fermi energy is it sensible to discuss discrete s.p. energies. 
For low-lying holes or high-lying particles the s.p. strength 
is strongly fragmented in realistic cases --- 
mainly due to short-range correlations~\cite{DiB2004}. 

\subsection{Transition strength} 

We consider the response of the nucleus to an external field described by the 
s.p. multipole operator 
\begin{equation} 
O_{{J}^{\pi}{M}} = \sum_{\alpha m_{\alpha}  \beta m_{\beta}} 
\langle \alpha m_{\alpha} | O_{{J}^{\pi}{M}} | \beta m_{\beta} \rangle 
  a_{\beta m_{\beta}}^{\dagger} a_{\alpha m_{\alpha}} 
. 
\label{E:spo} 
\end{equation} 
The corresponding strength distribution is given by 
\begin{equation} 
R(E) = 
\sum_{\nu} | \langle \nu | 
O_{{J}^{\pi}{M}} | 0 \rangle |^2 \delta (E - E_{\nu}) 
, 
\label{E:strd}  
\end{equation} 
where the sum runs over states with quantum numbers ${J}^{\pi}$. 
The transition matrix element between the ground state 
and the excited state $| \nu \rangle $  
[see Eq.~(\ref{E:nu})] is given by 
\begin{eqnarray}  
\lefteqn{ \langle \nu | 
O_{{J}^{\pi}{M}} | 0 \rangle 
= } \nonumber \\ 
&&
\sum_{\alpha m_{\alpha}  \beta m_{\beta}} 
\langle \alpha m_{\alpha} | O_{{J}^{\pi}{M}} | \beta m_{\beta} \rangle 
\nonumber \\ && \mbox{} \hspace{13mm}\times  
 \langle 0 | Q_{\nu, {J}^{\pi}{M}}  a_{\beta m_{\beta}}^{\dagger} a_{\alpha m_{\alpha}} 
| 0 \rangle 
. 
\end{eqnarray} 
After we substitute for the operator $Q$ using 
Eqs.~(\ref{E:nu}),~(\ref{E:Aop}), and (\ref{E:Nmat})  
we obtain 
\begin{eqnarray} 
\lefteqn{ \langle \nu | 
O_{{J}^{\pi}{M}} | 0 \rangle 
= } \nonumber \\ && 
\hat{J}^{-1}  
\sum_{ph} 
[ 
                    X_{ph}^{\nu , {J}^{\pi} \ast      } 
+ (-1)^{{J}} Y_{ph}^{\nu , {J}^{\pi} \ast      } 
] 
\nonumber \\ && \mbox{} \hspace{13mm} \times  
D_{ph}^{1/2} 
\langle p || O_{{J}} || h \rangle 
. 
\label{E:nu0me}
\end{eqnarray} 
The s.p. transition amplitudes are renormalized 
by a factor $D_{ph}^{1/2}$ due to the ground-state correlations. 

In the UCOM framework the operators of all observables must be 
transformed in a consistent way. In particular, the same 
unitary transformation as used for the nuclear Hamiltonian 
has to be applied to the transition operators as well. 
Thus, correlated transition operators $C^{\dagger} O_{{J}^{\pi}} C$ should be used in 
Eq.~(\ref{E:nu0me}). However, it has been shown~\cite{PPH2006} that  
the effect is negligible, at least for the monopole and quadrupole operators. 
 
Having solved the ERPA equations, 
one can use Eqs.~(\ref{E:strd}) and (\ref{E:nu0me}) to 
calculate the 
response of nuclei to external fields of interest. 
In this work we examine the response to isoscalar (IS) and isovector (IV) 
operators with natural parity, $\pi = (-1)^{{J}}$, 
of the usual form 
\begin{eqnarray}
O_{{J}^{\pi}{M}}^{\text{IS}} &=& 
\sum_{i=1}^{A} r_i^{{K}} 
Y_{{JM}}(\hat{r}_i) 
\\ 
O_{{J}^{\pi}{M}}^{\text{IV}} &=& 
\sum_{i=1}^{A} \tau_{z}^{(i)}  r_i^{{K}} 
Y_{{JM}}(\hat{r}_i) 
\end{eqnarray}
for 
${J} \neq 1$,  
with ${K}={J}$ for ${J}\geq 2$ 
and ${K}=2$ for ${J}=0$. 
For the dipole operators we will use 
the corresponding effective forms which remove the mixing with 
the spurious 
center-of-mass motion, 
\begin{eqnarray}
O_{1^-{M}}^{\text{IS}} &=& 
\sum_{i=1}^{A} (r_i^3 - \frac{5}{3}\langle r \rangle^2 r_i) 
Y_{1{M}}(\hat{r}_i) 
\\ 
O_{1^-{M}}^{\text{IV}} &=& 
 \frac{Z}{A}\sum_{i=1}^{N} r_i 
Y_{1{M}}(\hat{r}_i) 
-\frac{N}{A}\sum_{i=1}^{Z} r_i 
Y_{1{M}}(\hat{r}_i) 
. 
\end{eqnarray}
Thus, the spurious state 
is automatically eliminated from the strength distributions. 

Allowing for non-zero depletion (occupation) of the hole (particle) states 
in the ground state means that hole-hole and particle-particle transitions 
should be allowed, besides the $ph$ ones. An extension of the presented 
ERPA scheme has been introduced~\cite{GrC2000} to take these additional configurations 
into account. It was shown that, due to their omission, the energy-weighted sum rules are 
not preserved by the ERPA.  
For closed-shell nuclei, however, like those considered here, 
the deviation 
should be small.

\section{Results} 
\label{Sresults} 

We have applied the ERPA method to examine the response of closed-shell 
nuclei to isoscalar (IS) and isovector (IV), natural-parity external fields. 
We will present results obtained using the correlated Argonne V18 interaction with 
$I_{\vartheta}^{(1,0)} = 0.09$~fm$^3$. 
In most cases we have used a harmonic-oscillator basis of 13 shells, 
with an oscillator parameter $a_{\text{HO}}=1.8$~fm. 
We will call this set of parameters ``standard", for brevity. 
The term will mean also that both natural- and unnatural-parity modes 
(but not charge-exchange ones) of ${J}\leq J_{\max}=4$ have been taken into account 
in the evaluation of g.s. correlations.  
In certain instances we have 
changed the size of the basis, to 
test the convergence of our results, or used a different $J_{\max}$ value, 
or only natural-parity states. 

The main question is how our RPA results, such as those 
presented in Ref.~\cite{PPH2006}, will be affected by the ground-state 
correlations taken into account by the present scheme. 
It is instructive to 
first discuss the ground-state properties calculated within ERPA. 
Indeed, the s.p. occupation probabilities 
enter directly both the ERPA equations, 
Eqs.~(\ref{E:ERPAeq}),~(\ref{E:Amat}),~(\ref{E:Bmat}), and the calculation of the transition strength, 
Eq.~(\ref{E:nu0me}). They renormalize the 
matrix elements of the interaction in the former case and of the $ph$ transition 
operator in the latter, via the quantities $D_{ph}$.  
For given s.p. states, a more depleted Fermi sea would imply 
a weaker residual interaction and reduced $ph$ transition strength. 
Of course, the s.p. eigenstates are changed by the self-consistent solution 
of the ERPA equations and affect the final result as well.   

\subsection{Ground-state properties} 

For the nuclei \nuc{16}O, \nuc{40}{Ca}, \nuc{90}{Zr}, and \nuc{100}Sn 
and for the standard parameter set defined above 
we have evaluated the HF eigenenergies, the eigenenergies $e_{\alpha}$ 
of the new s.p. Hamiltonian and   
the centroid energies 
$\varepsilon_{\alpha}$ defined in Eqs.~(\ref{E:spvareh}) and (\ref{E:spvarep}). 
None of these sets of s.p. energies 
are corrected for the center-of-mass energy. 

\begin{figure} 
  \begin{center}  
  \includegraphics[height=10cm]{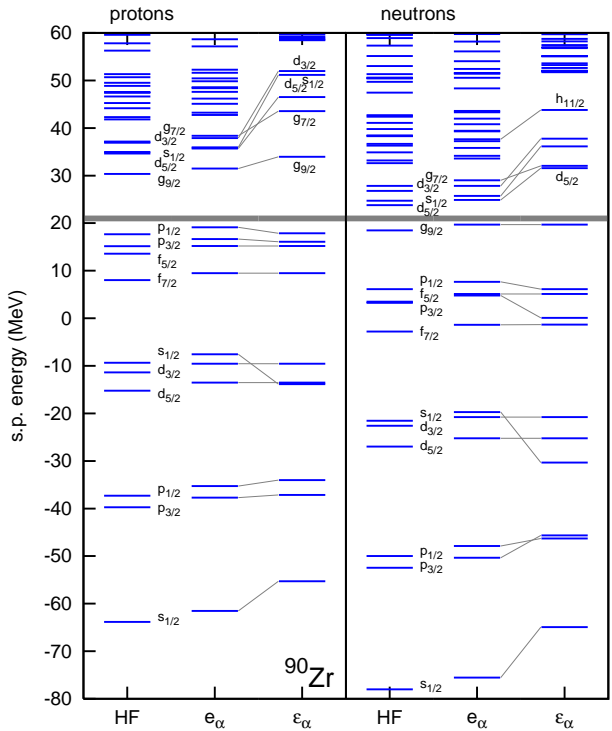} 
  \caption{%
  (Color online) 
  The Hartree-Fock eigenenergies, the eigenenergies $e_{\alpha}$ 
  of the new s.p. Hamiltonian and   
  the pick-up and removal centroid energies 
  $\pm\varepsilon_{\alpha}$, Eqs.~(\ref{E:spvareh}),~(\ref{E:spvarep}), 
  for the nucleus \nuc{90}{Zr}, evaluated using the 
  standard parameter set (see text). The energies are not 
  corrected for the center-of-mass energy. 
  A thick grey line separates the particle states from the hole states.  
  \label{F:speb}
  } 
  \end{center} 
  \end{figure}

In Fig.~\ref{F:speb} we show our results for the 
\nuc{90}{Zr} nucleus. The thick grey line separates the 
particle states from the hole states. 
One can observe that the spectrum of the new eigenenergies $e_{\alpha}$ is more dense 
than the HF one. The centroid energies $\varepsilon_p$ and $\varepsilon_h$ 
are even closer to each other, on average, while the Fermi gap in this case appears larger. 
Indeed, the lowest pickup centroid energy corresponds to the strength distribution 
to one or more particle eigenstates --- since 
there is no mixing between particle and hole configurations. 
Therefore, its value cannot be lower than the 
lowest particle state's $e_{\alpha}$. Similar arguments 
explain why $e_{\alpha}\geq\varepsilon_{\alpha}$ 
for the highest hole level and $e_{\alpha}\leq\varepsilon_{\alpha}$ for the 
most bound one. 
Let us note also that, in general, the ordering of the 
$\varepsilon_{\alpha}$ levels is different than for the $e_{\alpha}$ 
and HF ones. 
In total, the $ph$ energies 
$e_p-e_h$ 
are smaller than their HF 
counterparts which enter the RPA equations. 
This means that 
the ``unperturbed spectrum" corresponding to the ERPA equations 
(the one obtained by setting the $B$ matrix and the last term in Eq.~(\ref{E:Amat}) 
equal to zero, and assuming that the prefactor of the 
first term in Eq.~(\ref{E:Amat}) is roughly equal to one) 
will be denser than the unperturbed HF spectrum. 
The above remarks hold for all examined nuclei.  

\begin{figure} 
  \begin{center}  
  \includegraphics[height=6.8cm]{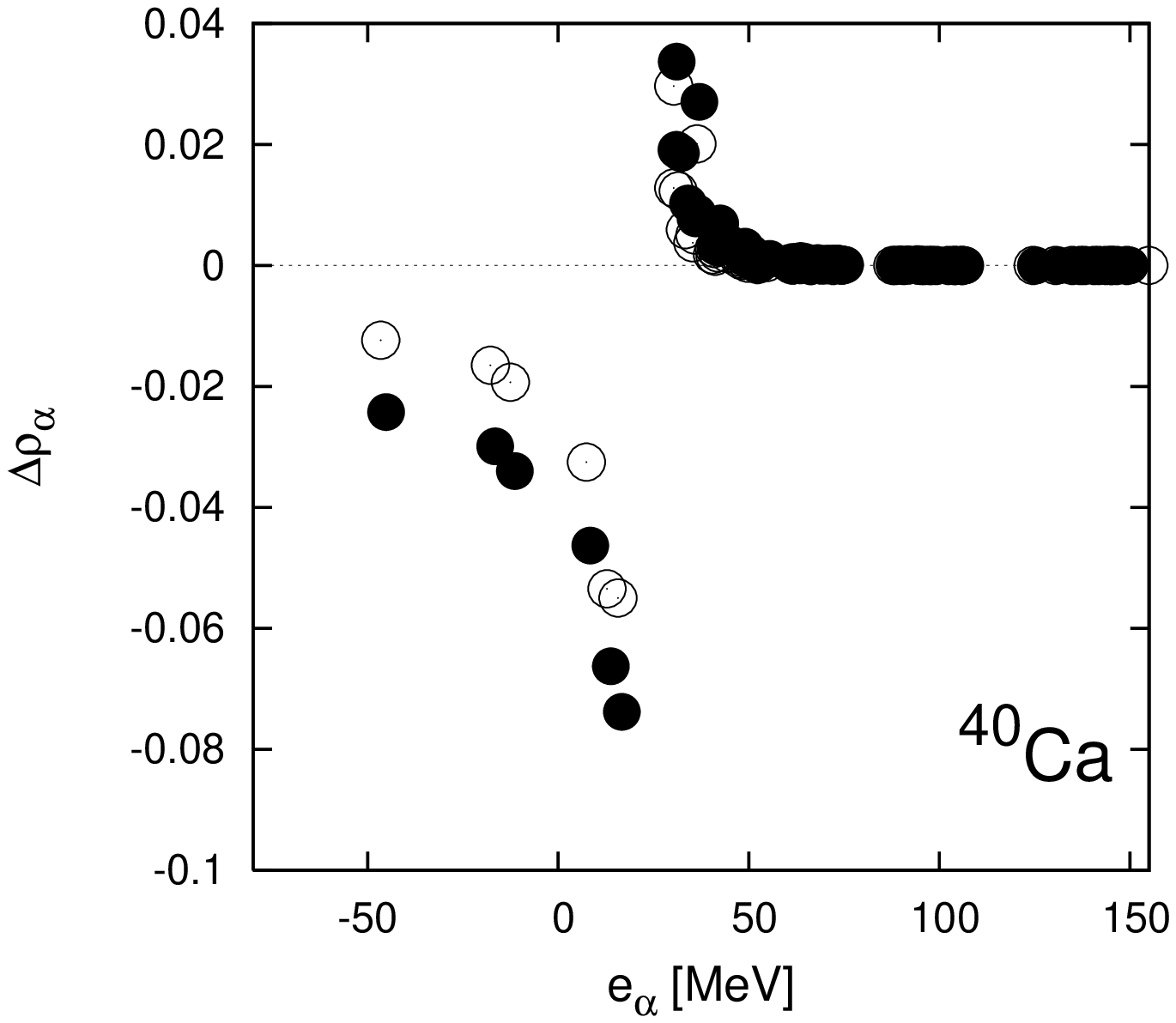} 
  \includegraphics[height=6.8cm]{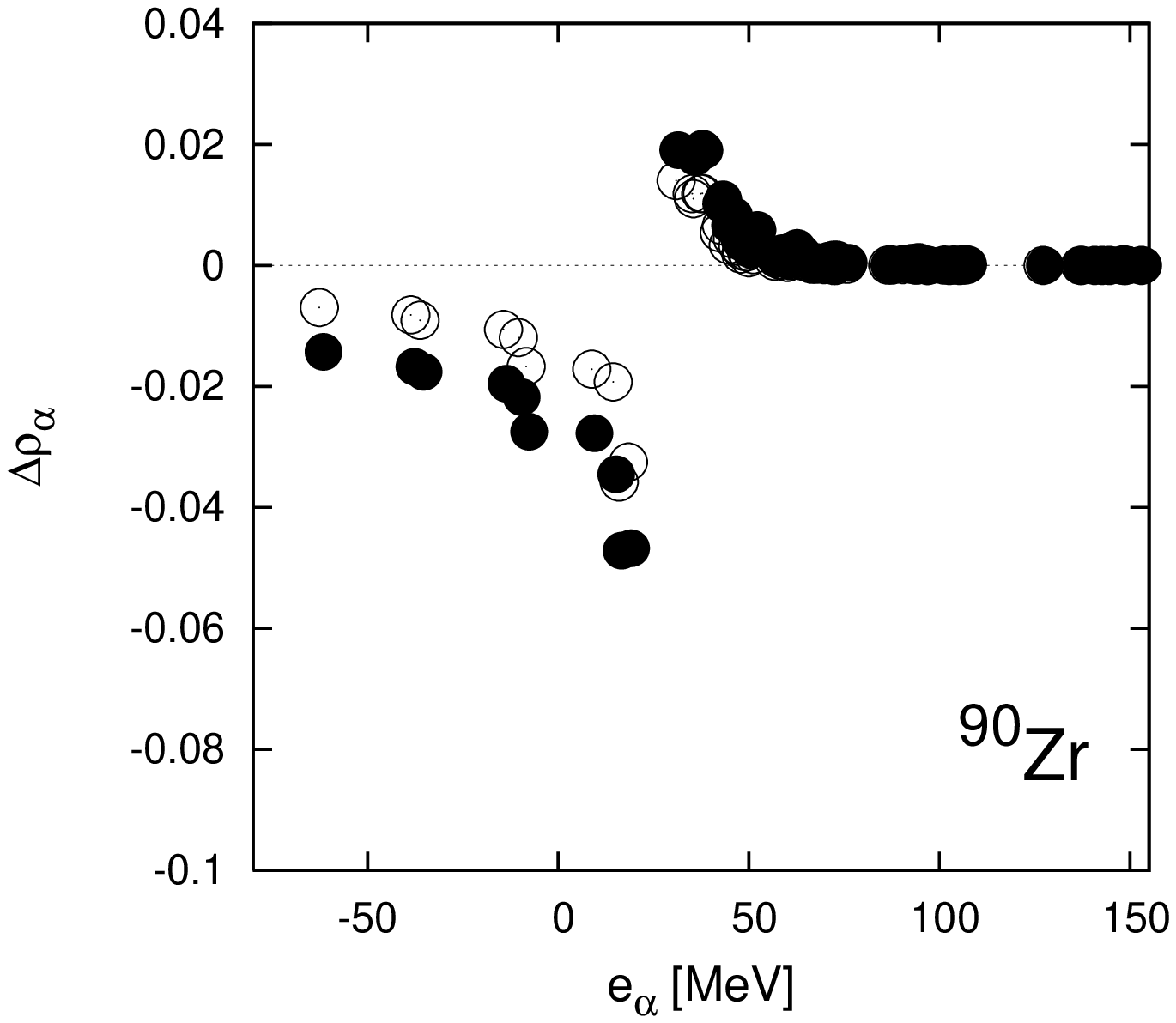} 
  \caption{%
  The deviation 
  $\Delta\rho_{\alpha} = \rho_{\alpha}-\rho_{\alpha}^{(0)}$ 
  of the occupation probabilities $\rho_{\alpha}$  
  from their HF values $\rho_{\alpha}^{(0)}$  
  is shown for the 
  proton single-particle levels of the nuclei  
  \nuc{40}{Ca} and \nuc{90}{Zr}. 
  Filled circles: standard parameter set; open circles: only natural-parity 
  states taken into account.  
  The energies are not corrected for the center-of-mass motion. 
  \label{F:ocp} 
  }
  \end{center} 
  \end{figure}

In Fig.~\ref{F:ocp} the deviation 
$\Delta\rho_{\alpha} = \rho_{\alpha}-\rho_{\alpha}^{(0)}$ 
of the occupation probabilities $\rho_{\alpha}$  
from their HF values $\rho_{\alpha}^{(0)}$  
(1 for holes, 0 for particles) is shown for the 
proton s.p. levels of the nuclei  
\nuc{40}{Ca} and \nuc{90}{Zr}. The full circles correspond to the 
standard parameter set. In the case of the open circles, 
only natural-parity excitations have been taken into account. 
More than half the total depletion seems to be accounted for 
by these excitations. 

In Table~\ref{T:sigma} we list the values 
  \begin{table}[b] 
  \caption{ 
  The values of the mean square deviation per particle $\sigma$, Eq.~(\ref{E:sigmaNO}), 
  found when all ${J}^{\pm}$ excitations ($\sigma_{\text{NUP}}$ - standard parameter set) 
  and only natural-parity ($\sigma_{\text{NP}}$) 
  ones are considered. In square brackets, for \nuc{90}{Zr}, results obtained within a larger s.p. basis 
  (15 shells) are shown.  
  \label{T:sigma} 
  }
  \begin{ruledtabular} 
  \begin{tabular}{lcccc} 
                                      &  \nuc{16}{O}  &  \nuc{40}{Ca}  &  \nuc{90}{Zr}  &  \nuc{100}{Sn} \\ 
   \hline 
   $\sigma_{\text{NUP}}\times 10^3$ &    3.43       &    2.23        &    1.11[1.17]  &    1.37        \\ 
   $\sigma_{\text{NP}} \times 10^3$ &    1.85       &    1.05        &    0.46[0.48]  &    0.57        \\ 
  \end{tabular}  
  \end{ruledtabular}  
  \end{table} 
of the mean square deviation per particle $\sigma$, 
Eq.~(\ref{E:sigmaNO}), 
that we find for the four nuclei 
\nuc{16}{O}, \nuc{40}{Ca}, \nuc{90}Zr, \nuc{100}{Sn} 
when all ${J}^{\pm}$ excitations 
($\sigma_{\text{NUP}}$ - standard parameter set) 
or only natural-parity ones ($\sigma_{\text{NP}}$) 
are considered. (Note that 
no charge-exchange excitations are considered.) 
$\sigma$ characterizes the depletion of the Fermi sea 
due to correlations. 
In square brackets, for \nuc{90}{Zr}, results obtained within a larger s.p. basis 
(15 shells) are shown, to verify that 
our conclusions remain valid.  
The ratio $\sigma_{\text{NP}}:\sigma_{\text{NUP}}$ 
is roughly 50\%.  
We notice that the values of $\sigma$ are of the order 
$10^{-3}$, reflecting the effect of 
long-range correlations. 
As mentioned in Sec.~\ref{S:gsp}, 
such $\sigma$-values are typical of models which do not 
take into account short-range correlations. 
In principle, these are dealt with by the UCOM, but a full 
treatment to this end has not been pursued here. Indeed, 
when calculating the OBDM and the occupation probabilities 
$\rho_{\alpha}$, we have used the uncorrelated operators  
$a^{\dagger}_{\alpha} a_{\alpha '}$. 
For a consistent UCOM treatment, however,  
correlated operators $C^{\dagger} a^{\dagger}_{\alpha} a_{\alpha '} C$ 
should be considered. Such calculations are not trivial and 
lie beyond the scope of the present work. As a relevant example where the 
potential of the UCOM was fully exploited in describing short-range 
correlations in nuclei, we mention 
the calculations of Ref.~\cite{NeF2003}, where the high-momentum tail of the 
s.p. momentum distribution 
is reproduced. 

For \nuc{40}{Ca} and \nuc{90}{Zr} 
we can compare our present results for the depletion of 
the Fermi sea to the ones obtained within 
second-order perturbation theory (PT) and reported in Ref.~\cite{RPP2006}. 
In general the occupation probabilities 
of hole states are larger in the case of ERPA, especially for \nuc{90}{Zr}. 
We note that within PT more configurations are taken into account in the 
evaluation of ground-state correlations, namely all the values of ${J}^{\pm}$ 
available within the basis, including charge-exchange excitations. 
More importantly, 
the use of non-renormalized PT is known to overestimate the depletion 
of the Fermi sea~\cite{VWV1992}. 
This can explain also why for \nuc{90}{Zr} the particle occupation 
probabilities obtained within ERPA are much smaller. 
For \nuc{40}{Ca} they are larger only close to the Fermi energy. 
For both nuclei 
the ERPA values of the particle occupation probabilities drop much faster to zero as the energy increases, 
compared to PT. The 
use of the natural-orbital basis, rather than the HF one, is responsible for this behaviour;  
it is not a matter of RPA vs. PT --- cf. the particle occupation 
probabilities reported in Ref.~\cite{LeW1990} within (S)RPA in the HF basis.

\subsection{Collective excitations}  

\begin{figure*} 
  \begin{center}  
  \includegraphics[height=5.3cm,angle=270]{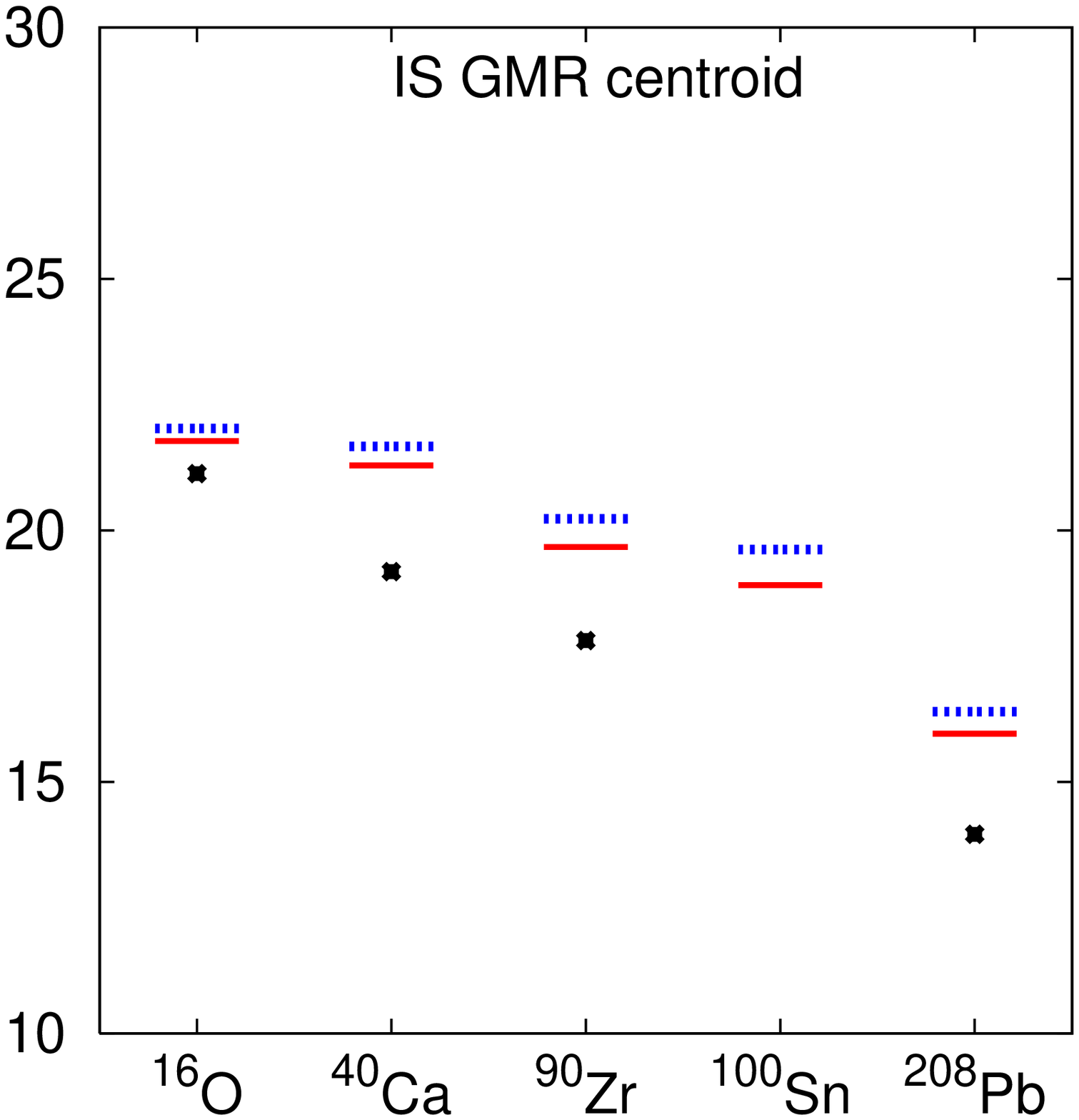} 
  \includegraphics[height=5.3cm,angle=270]{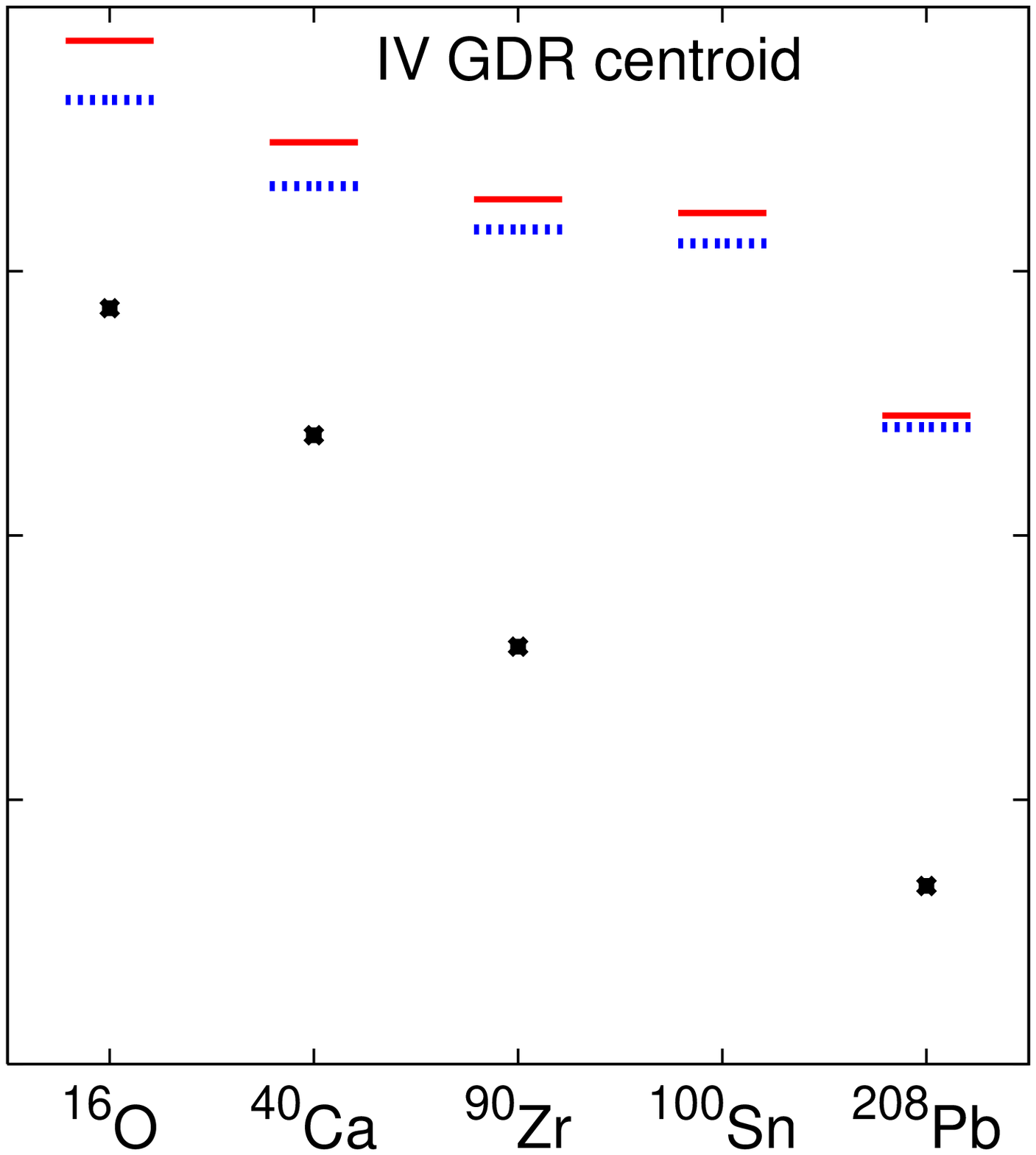} 
  \includegraphics[height=5.3cm,angle=270]{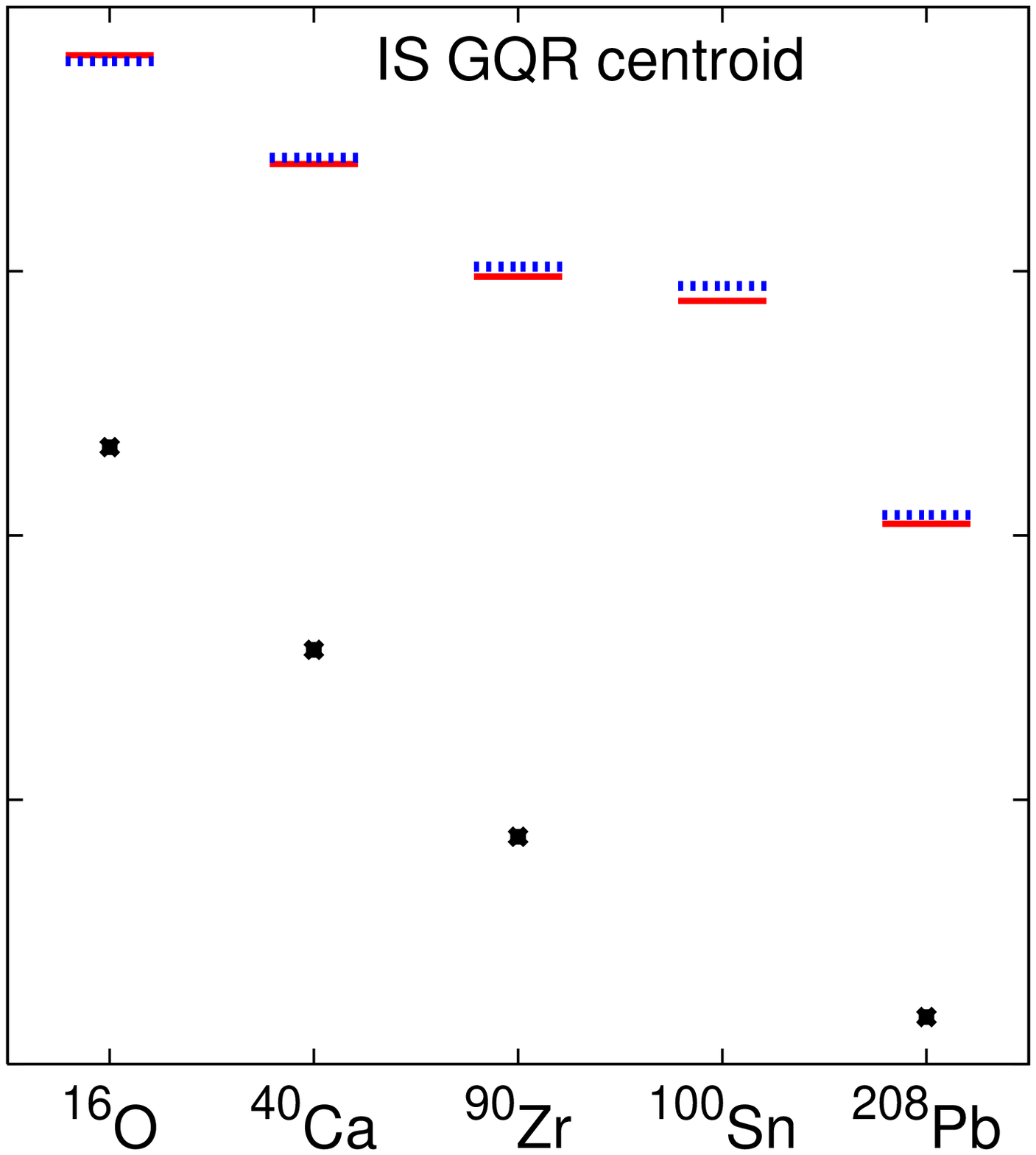} 
  \caption{%
  (Color online)  
  Centroid energies of the IS GMR, IV GDR, and IS GQR 
  obtained for the nuclei 
  \nuc{16}{O}, \nuc{40}{Ca}, \nuc{90}{Zr}, \nuc{100}{Sn}, and \nuc{208}{Pb} 
  within the RPA (red solid bars) and ERPA (blue dotted bars), compared with 
  experimental data (black points --- for references see text). 
  The standard parameter set was used, except for \nuc{208}{Pb}, 
  where $J_{\max}=3$ and a s.p. basis of 15 shells were employed. 
  In calculating the centroids, the strength up to 50~MeV 
  was taken into account.   
  \label{F:centr} 
  }
  \end{center} 
  \end{figure*}

\begin{figure} 
  \begin{center}  
  \includegraphics[height=5.3cm,angle=270]{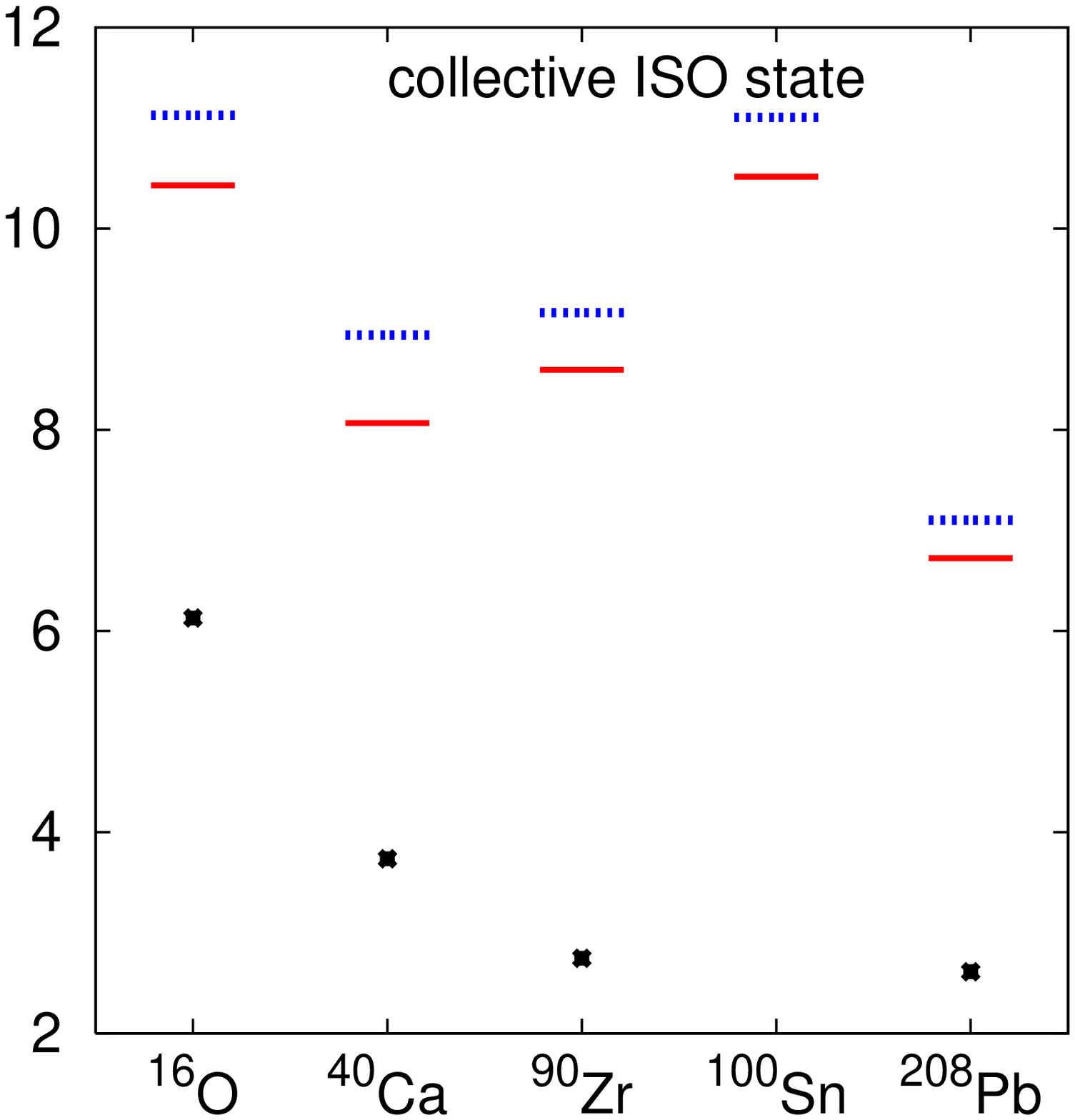} 
  \caption{%
  (Color online)  
  As in Fig.~\ref{F:centr}, energy of the low-lying IS octupole (ISO) state.  
  \label{F:centr3} 
  }
  \end{center} 
  \end{figure}

\begin{figure*} 
  \begin{center}  
  \includegraphics[height=7.3cm,angle=270]{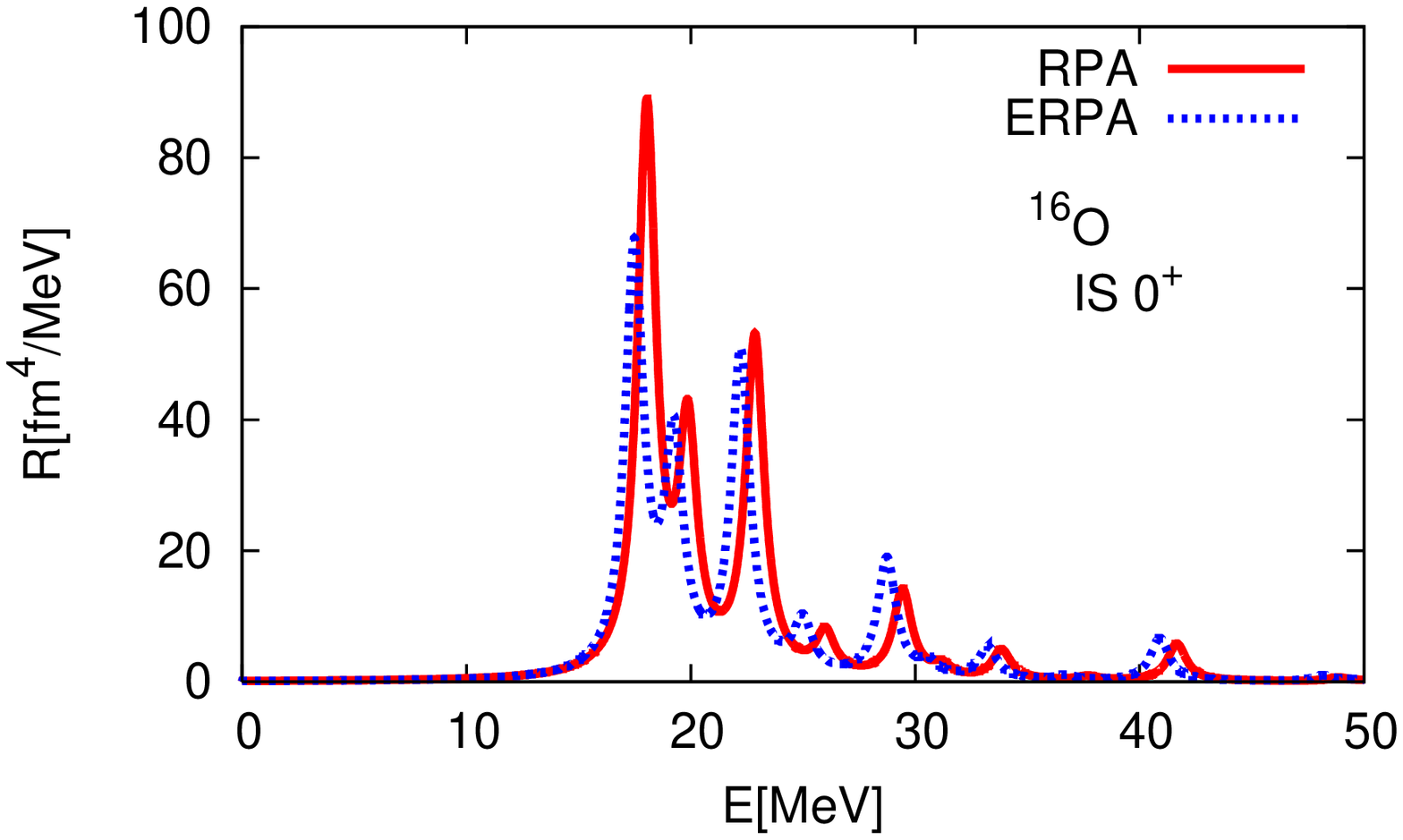} 
  \hspace{2mm} 
  \includegraphics[height=7.3cm,angle=270]{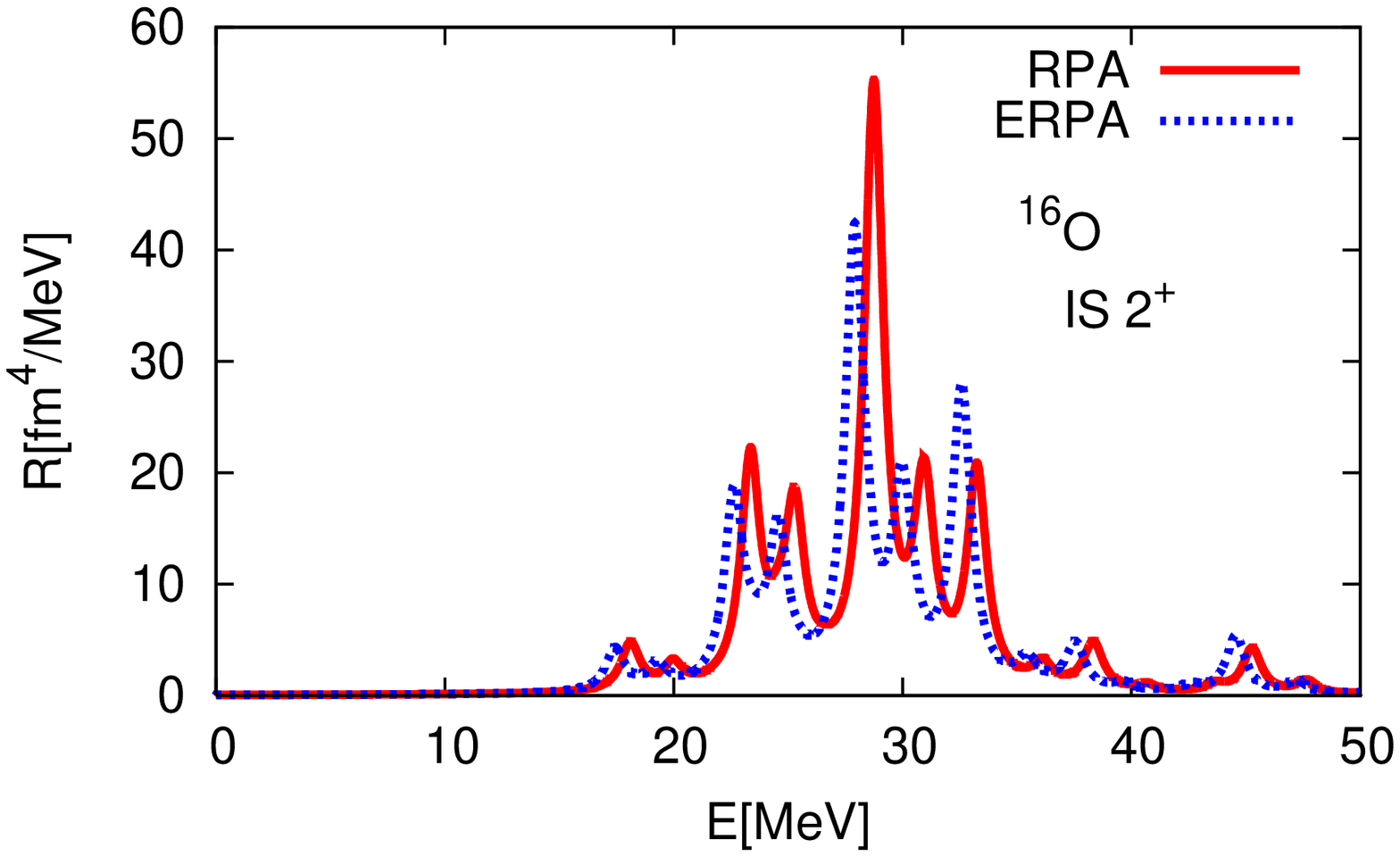} 
  \\ 
  \includegraphics[height=7.3cm,angle=270]{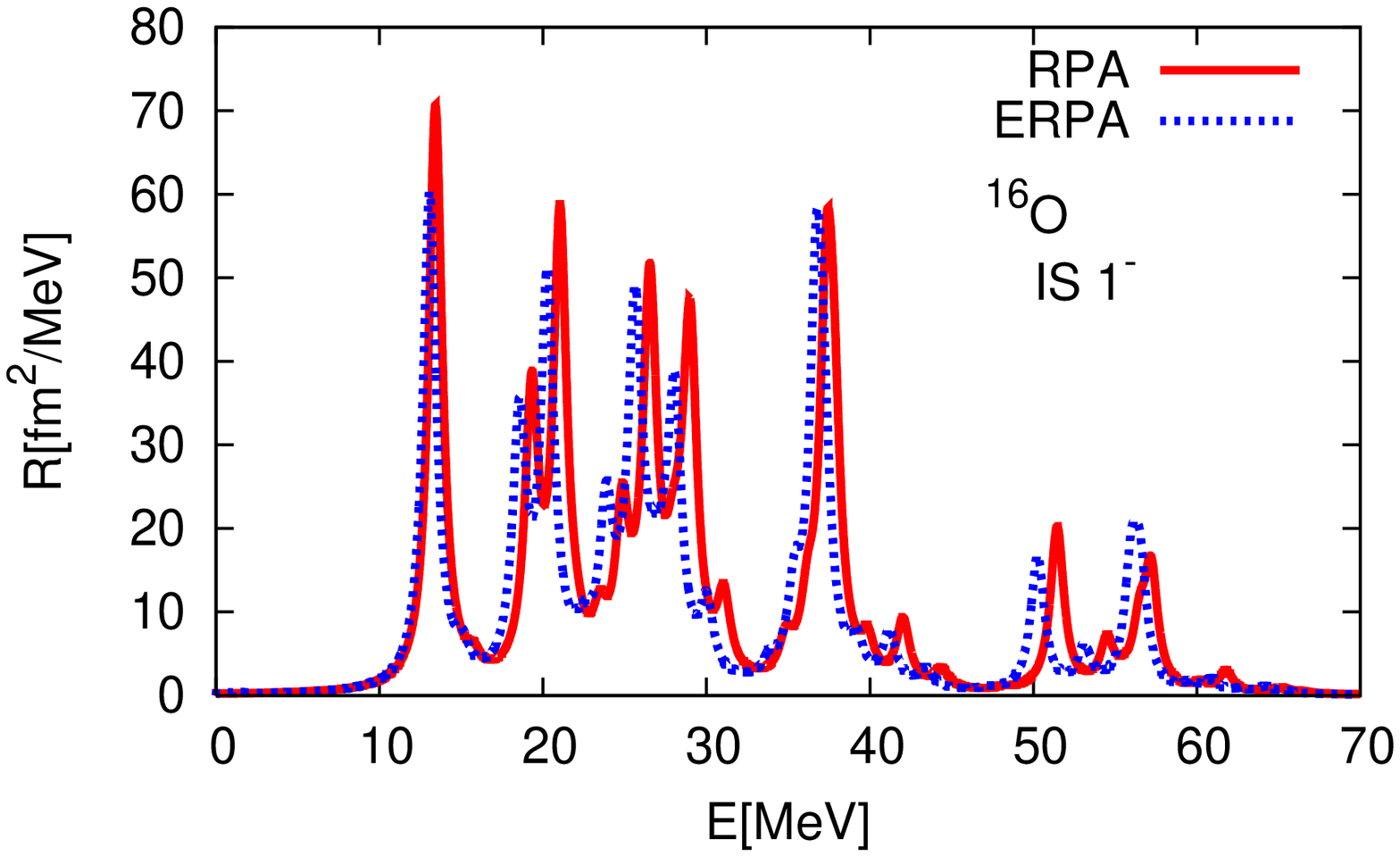} 
  \hspace{2mm} 
  \includegraphics[height=7.3cm,angle=270]{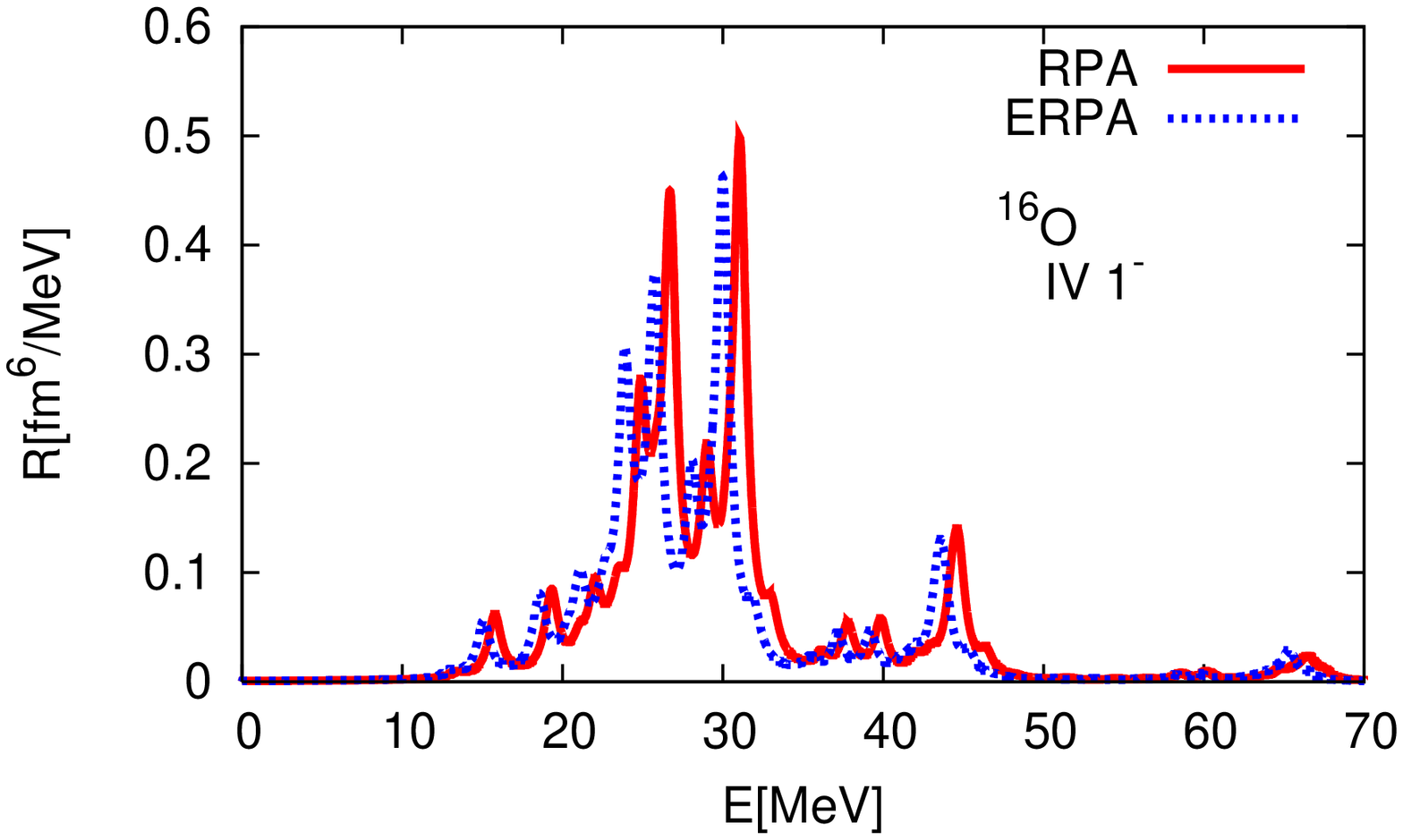} 
  \caption{%
  (Color online)  
  IS monopole, IS quadrupole, and  
  IS and IV dipole 
  strength distributions of \nuc{16}{O} 
  computed within RPA (red solid lines) and ERPA (blue dotted lines). 
  The discrete strength distributions have been folded with a 
  Lorentzian of width $\Gamma =2$~MeV. 
  The standard parameter set was used. 
  \label{F:ivdo} 
  }
  \end{center} 
  \end{figure*}

\begin{figure*} 
  \begin{center}  
  \includegraphics[height=7.3cm,angle=270]{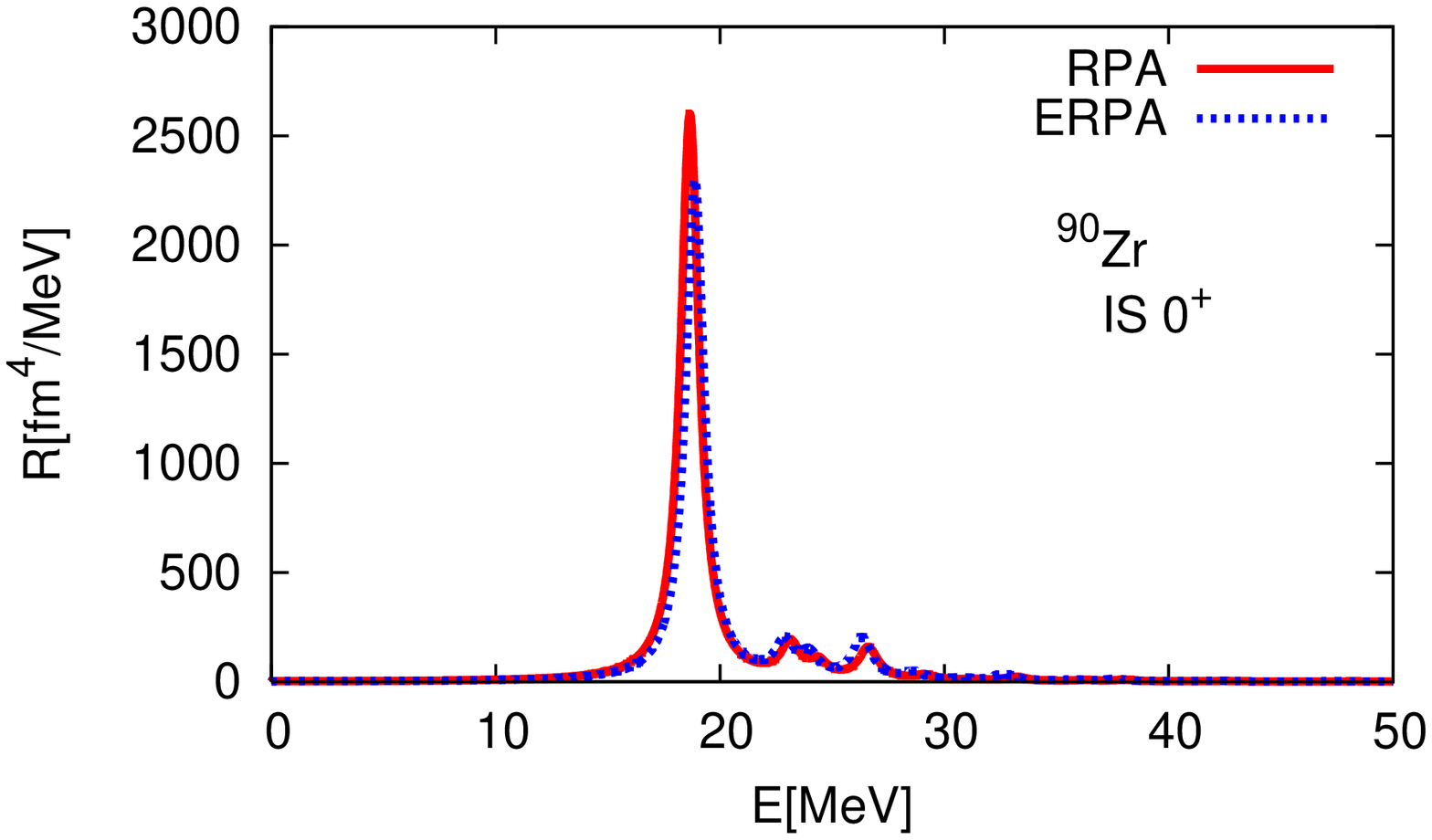} 
  \hspace{2mm} 
  \includegraphics[height=7.3cm,angle=270]{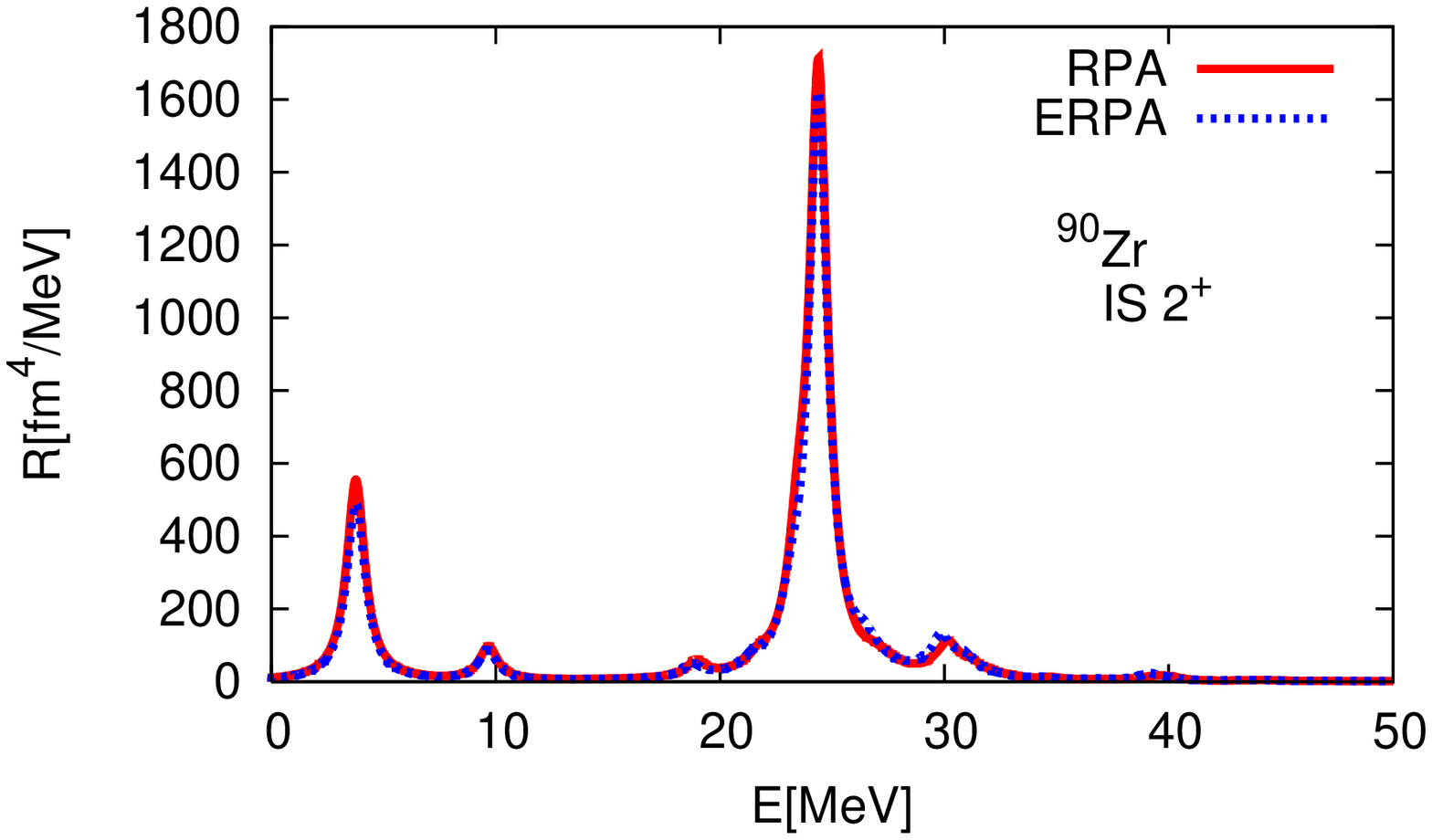} 
  \\ 
  \includegraphics[height=7.3cm,angle=270]{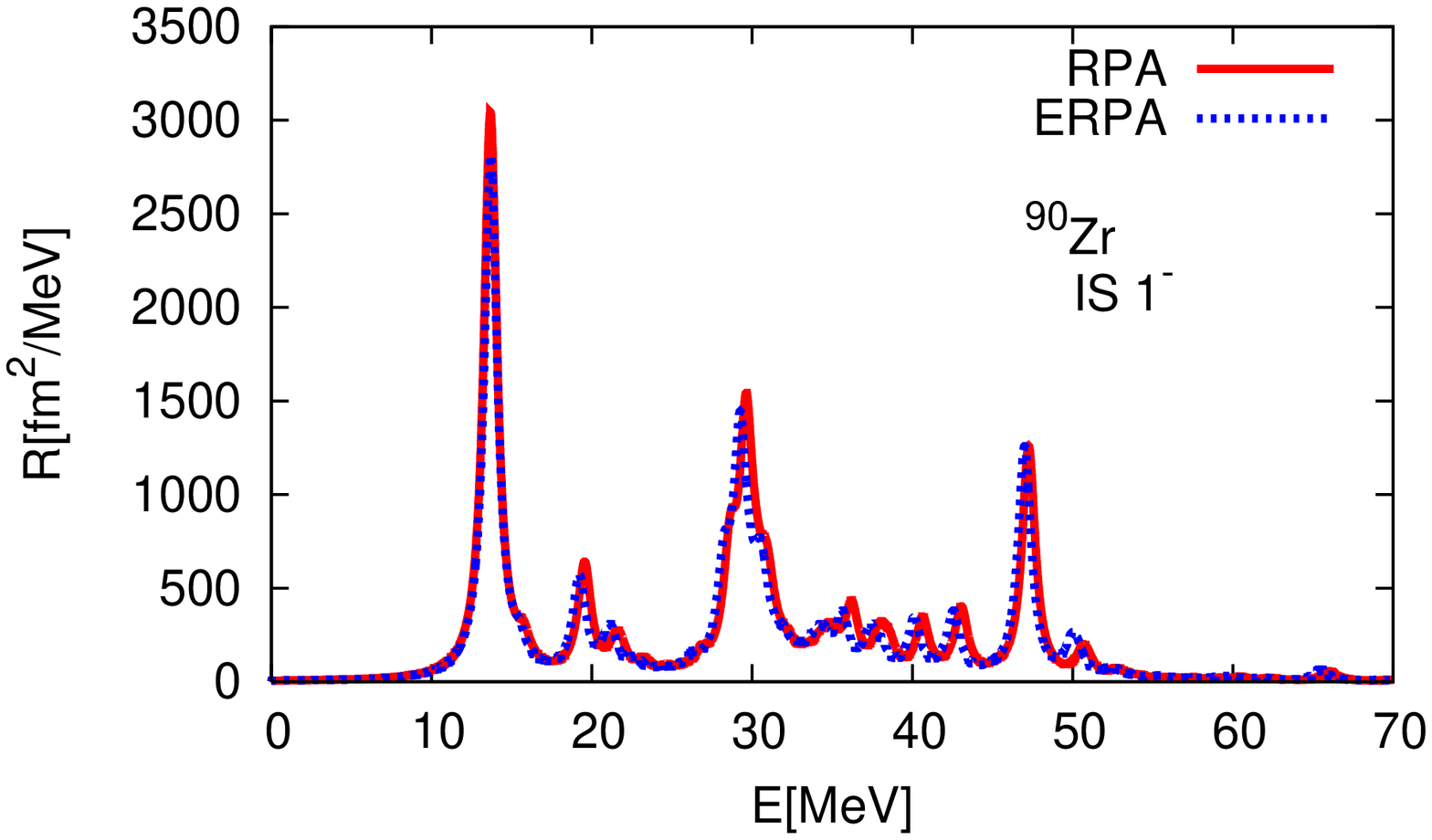} 
  \hspace{2mm} 
  \includegraphics[height=7.3cm,angle=270]{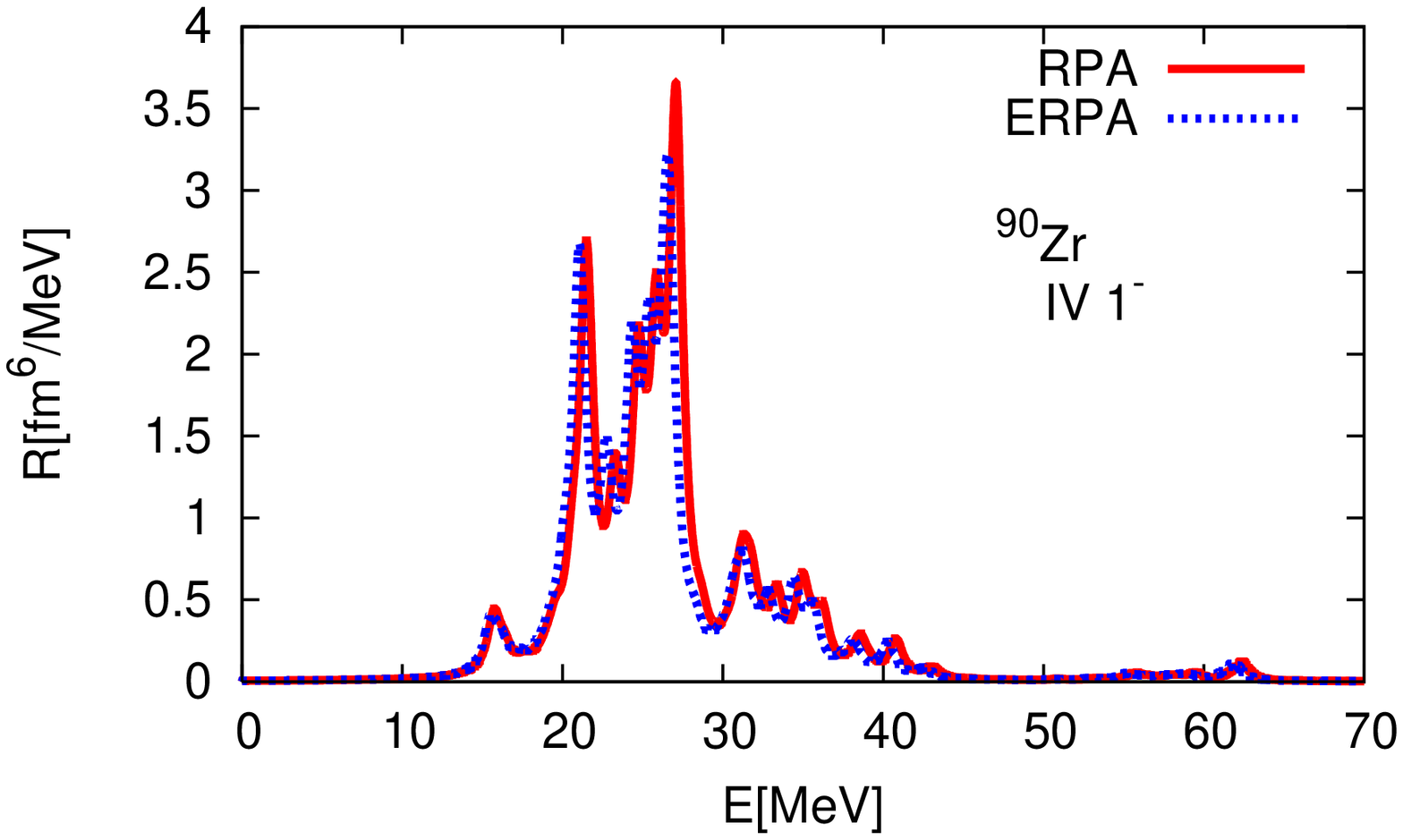} 
  \caption{%
  (Color online)  
  As in Fig.~\ref{F:ivdo}, for \nuc{90}{Zr}. 
  \label{F:ivdzr} 
  }
  \end{center} 
  \end{figure*}

In Fig.~\ref{F:centr} 
we show the centroid energies of the  
IS GMR, IV GDR, and IS GQR 
obtained for the nuclei 
\nuc{16}{O}, \nuc{40}{Ca}, \nuc{90}{Zr}, \nuc{100}{Sn}, and \nuc{208}{Pb} 
within the RPA (solid bars) and ERPA (dotted bars), compared with 
some experimental data (points). 
In Fig.~\ref{F:centr3} 
we show the energy of the low-lying IS octupole (ISO) state.  
As examples, we show in Figs. \ref{F:ivdo} 
and \ref{F:ivdzr} 
the corresponding IS monopole, IV dipole, and IS quadrupole strength 
distributions for \nuc{16}{O} and \nuc{90}{Zr}. 
Also shown in these figures are the 
IS dipole strength distributions. 
In all cases shown in Figs.~\ref{F:centr} -- \ref{F:ivdzr} 
the standard parameter set was used, except for \nuc{208}{Pb}, 
where a basis of 15 shells was employed to ensure better convergence in this 
heavy nucleus and $J_{\max}=3$ was used to avoid an extremely time-consuming 
calculation. 
The strength distributions up to 50~MeV 
were taken into account 
when calculating the centroid energies, 
which are defined as the first energy moment of the distribution, $m_1$, 
divided by the total strength, $m_0$. 
The experimental centroids of the IS GMR and the IS GQR were taken from 
Refs. \cite{LCY2001} (\nuc{16}{O}), \cite{YLC2001} (\nuc{40}{Ca}), 
\cite{You2004a} (\nuc{90}{Zr}), and \cite{You2004b} (\nuc{208}{Pb}). 
Photoabsorption cross sections were found in 
Refs. \cite{LNH1987} (\nuc{16}{O}), \cite{Vey1974} (\nuc{40}{Ca}), 
\cite{Ber1967} (\nuc{90}{Zr}), and \cite{Vey1970,BeF1975} (\nuc{208}{Pb}) 
and the centroids $m_1/m_0$ of 
the corresponding IV GDR strength distributions were evaluated from those. 
Experimental energies of the collective octupole state were 
adopted from Ref.~\cite{KiS2002}.

From Fig.~\ref{F:centr} it is evident that the energy of the GDR 
drops when ground-state correlations are explicitly considered within the ERPA, 
but not enough to reach the experimental data. In general, a decrease of 
no more than 1~MeV was achieved for the heavier nuclei. 
The same holds for the IV $0^+$ and $2^+$ centroids (not shown).  
The IS giant resonances, namely the IS GMR, GDR, and GQR, were less affected. 
In most cases, their energies were higher when evaluated within ERPA than 
within RPA. Apparently, in the IS case the weakening of the 
(attractive) residual interaction due to the depleted Fermi sea 
compensates for the compression of the s.p. spectra. 

The energy of the collective IS octupole state, Fig.~\ref{F:centr3}, 
increases by up to 1~MeV  
and the disagreement with the experimental data is worsened. 
We found that its strength within ERPA decreases with respect to RPA by 10-20\%. 
Still, the ERPA values of the strength are larger than the experimental ones by a factor 
1.6-2.2 for the nuclei considered. 
The energy of the low-lying IS quadrupole state for the nuclei 
\nuc{90}{Zr}, \nuc{100}{Sn}, and \nuc{208}{Pb} (not shown) 
increases due to the correlations by no more than 0.3~MeV 
and its strength decreases by 10-20\%.   

We found that, for all nuclei and excitation fields examined, 
the total strength $m_0$ and total energy-weighted strength $m_1$ 
decrease within ERPA. 
The relative change 
remains below 10\% in almost all cases, being largest for the lightest nuclei. 
Within ERPA the spurious dipole state is found at approximately 
1~MeV, i.e., somewhat further from zero than within RPA.  

\begin{figure*} 
  \begin{center}  
  \includegraphics[height=6.4cm,angle=270]{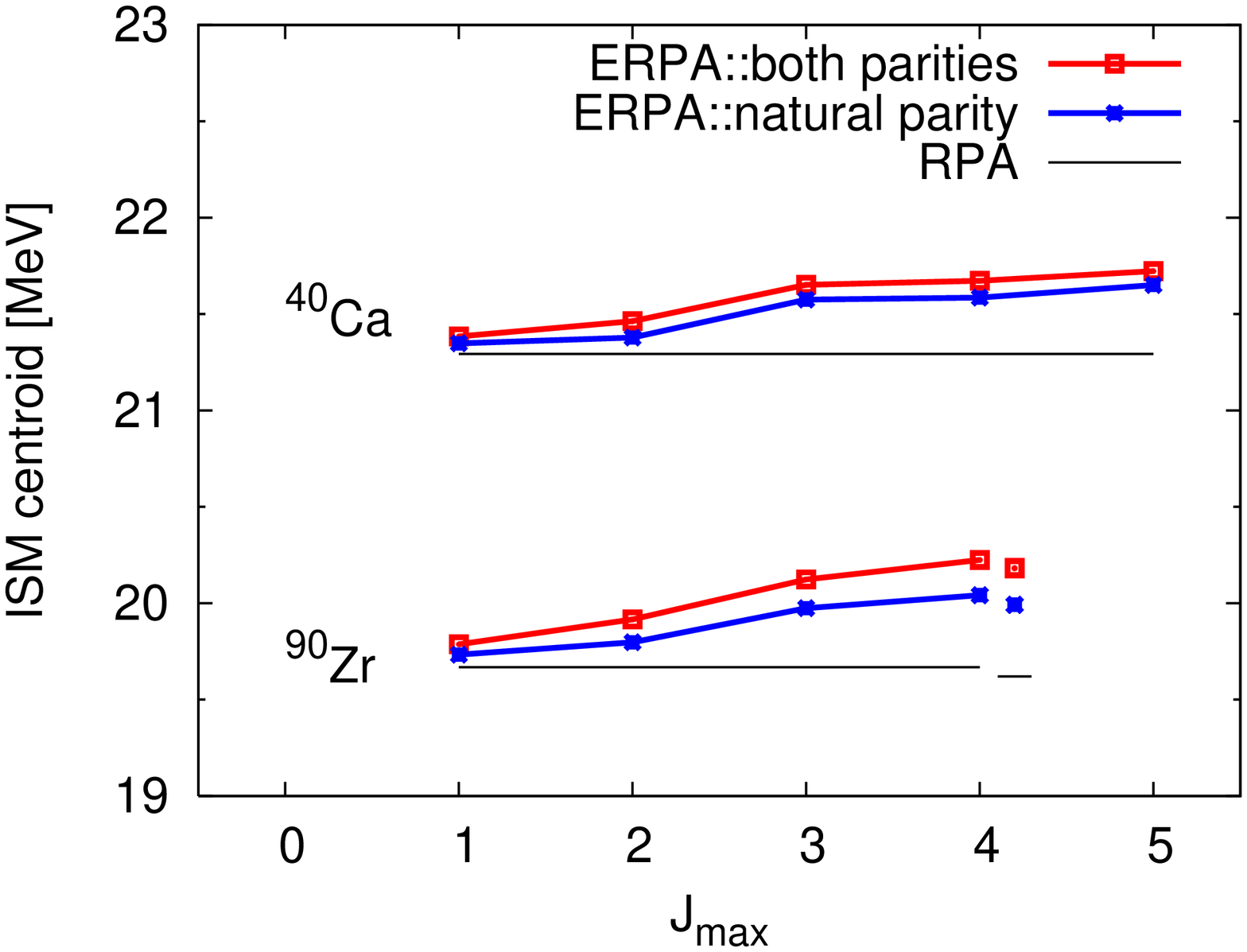} 
  \hspace{8mm} 
  \includegraphics[height=6.4cm,angle=270]{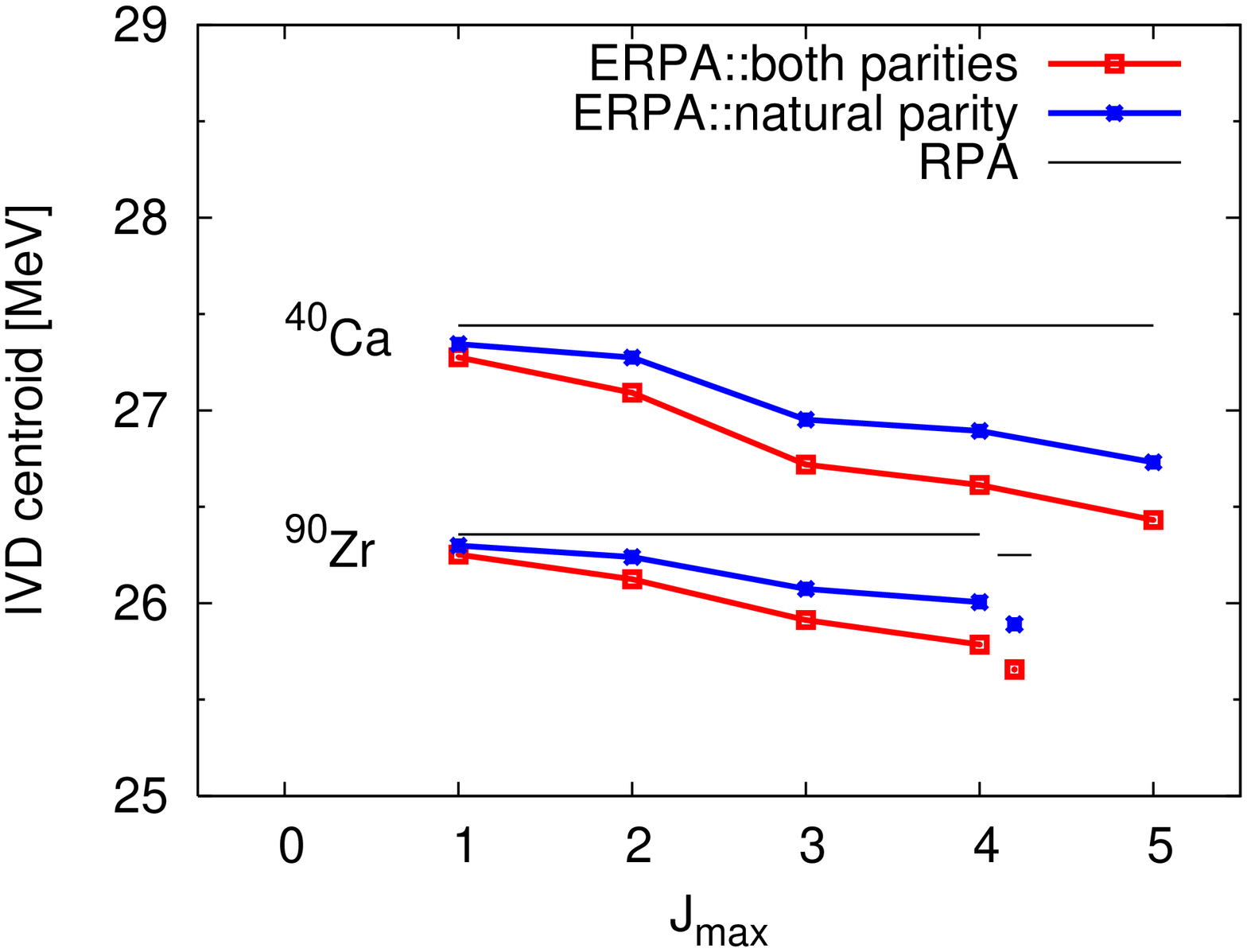} 
  \caption{%
  (Color online)  
  The cumulative effect 
  of allowing zero-point fluctuations of larger multipolarity $J_{\max}$ 
  on the IS GMR and IV GDR centroids of 
  \nuc{40}Ca and \nuc{90}{Zr}. 
  Blue crosses: using only natural-parity states. 
  Red squares: both natural- and unnatural-parity states were used. 
  A s.p. basis of 13 shells was used.
  Lines are drawn to guide the eye. 
  The horizontal lines show the RPA results for comparison. 
  For \nuc{90}{Zr} and $J_{\max}=4$ an extra set of data is shown in each panel, 
  detached from the lines. 
  These were obtained using a larger s.p. basis of 15 shells.  
  \label{F:ismivd} 
  }
  \end{center} 
  \end{figure*}

In Fig.~\ref{F:ismivd} 
we show, for \nuc{40}{Ca} and \nuc{90}{Zr}, 
the cumulative effect on the IS GMR and IV GDR centroids 
of allowing zero-point fluctuations of ever larger multipolarity $J_{\max}$. 
The crosses mark results obtained using only natural-parity states. 
The squares mark results obtained using both natural- and unnatural-parity states. 
Lines are drawn to guide the eye. 
The horizontal lines show the RPA results for comparison. 
In calculating the centroids, the corresponding strength distributions up to 50~MeV were 
taken into account. 
For \nuc{90}{Zr} and $J_{\max}=4$ an extra set of data is shown in each panel, 
detached from the lines. 
These were obtained using a larger s.p. basis of 15 shells 
and verify that our calculations have converged to a rather satisfactory degree 
with respect to the basis. 

In all cases a steeper change is observed between $J_{\max}=2$ and $3$, indicating the important effect of the 
low-lying collective $3^-$ state.  
We see also that the convergence of our results as $J_{\max}$ increases is rather slow. 
As RPA calculations of ground-state energies suggest~\cite{BPR2006}, 
in order to achieve good convergence when describing RPA correlations  
one may need to take into account $J_{\max}$ values up to approximately 10. 
For ERPA, such large values are beyond our computational capabilities at present.  
Since, however, no strongly collective states with multipolarities larger than 4 are expected, 
and based also on the results of Ref.~\cite{BPR2006}, 
it is reasonable to speculate that the difference between the RPA and ERPA 
results would be no more than doubled 
if higher values of $J_{\max}$ 
were considered. 
Of course, no charge-exchange excitations have been taken into account here. 
These can introduce additional correlations of non-negligible amplitude, 
albeit smaller than the non-exchange ones~\cite{BPR2006}.

\section{Summary and perspectives} 
\label{Ssummary} 

A correlated realistic interaction, derived from the 
Argonne V18 nucleon-nucleon interaction in the UCOM framework, 
was employed 
in calculations within the so-called Extended RPA (ERPA). 
The aim was to investigate 
to which extent 
such an interaction 
can describe nuclear collective motion in the framework of first-order RPA. 
Renormalized formulations of the RPA, 
such as the ERPA,  
consider explicitly 
the depletion of the Fermi sea in the 
ground state due to RPA correlations. 
In a previous work we used the same interaction within a 
RPA model~\cite{PPH2006}, 
where excitations are 
built on top of the uncorrelated HF ground state. 
The ERPA employed here allowed us 
to examine the effect of ground-state correlations 
on the 
excitation spectra. 
It was found that the effect on the properties of 
giant resonances is rather small. 
Compared to the standard RPA model, their centroid energies decrease by 
up to 1 MeV in the 
isovector channel. The isoscalar response is less affected in general. 
Ground-state properties obtained within the ERPA 
were compared with corresponding HF and perturbation-theory results and  
discussed as well.  

Our results with the correlated Argonne V18 
show that the RPA based on the uncorrelated HF ground state 
is a fairly good approximation to the 
actual RPA problem in the case of giant resonances,   
although with this interaction the HF solution 
is a bad approximation to the RPA ground state when one considers the nuclear binding energies. 
It should be noted that 
a theoretical error of 1~MeV in the centroid energies 
(or more, if charge-exchange and higher-mutipole correlations are included) 
can be important in certain applications. 
That said, the inability of our (E)RPA calculations 
to reproduce the experimental data should 
originate elsewhere. Up to now we have assumed that residual three-body 
forces can be neglected, based upon the fact that they contribute only marginally  
to the ground-state energy as calculated within many-body perturbation theory~\cite{RHP2005}. 
This is not necessarily a valid assumption. We are currently constructing 
a simple phenomenological zero-range three-body force, to be used along with 
the correlated two-nucleon interaction in future calculations. Preliminary results show that 
by using such a three-body force it is possible to improve on the description of observables such 
as nuclear radii and resonance energies while retaining the good reproduction of the experimental binding energies. 

Another important issue is that, within the RPA, one neglects the coupling to higher-order configurations 
($2p2h$ and beyond). 
Collective excitations can be significantly affected by 
the inclusion of $2p2h$ configurations, as preliminary SRPA results show. 
Therefore, SRPA calculations will also be the subject of future work. 
In principle, it is possible to combine the SRPA with a correlated ground state~\cite{DNS1990,GGC2006} 
for a most complete theoretical treatment of nuclear excitations.  

\begin{acknowledgments} 
This work was supported by the Deutsche Forschungsgemeinschaft 
within the SFB 634. 
\end{acknowledgments}

\end{document}